\documentclass[conference,compsoc]{IEEEtran}
\ifCLASSOPTIONcompsoc
      \usepackage[nocompress]{cite}
\else
    \usepackage{cite}
\fi
\ifCLASSINFOpdf
            \else
                  \fi

\usepackage{amsmath}
\usepackage{xcolor}
\usepackage{algorithmic}
\usepackage{graphicx}
\usepackage{xspace}
\usepackage{amssymb,epsfig,amsthm}
\usepackage{comment}
\usepackage{caption}
\usepackage{subcaption}
\usepackage{soul}
\usepackage{eucal}
\usepackage{multirow}
\usepackage{balance}
\usepackage{hhline}
\usepackage{url}
\usepackage{enumitem}
\usepackage{hyperref}
\usepackage{booktabs}
\usepackage{fancyhdr}
\newcommand{\system}{{\sc MANTIS}\xspace}

\newcommand\todo[1]{{\color{red} TODO: #1}}
\newcommand\eat[1]{}
\newcommand\fatih[1]{{\color{black}#1}}
\newcommand\fatihh[1]{{\color{black}#1}}
\newcommand\nabeel[1]{{\color{purple} N: #1}}
\newcommand\ik[1]{{\color{green} IK: #1}}

\newcommand\revision[1]{{\color{black} #1}}
\newcommand\repe[1]{{\color{black} #1}}
\newcommand\review[1]{{\color{black} #1}}
\newcommand\shorten[1]{{\color{black} #1}}
\newcommand\ccsupdate[1]{{\color{black} #1}}
\newcommand\spupdate[1]{{\color{black} #1}}
\newcommand\spfinal[1]{{\color{black} #1}}
\newcommand\eatcrv[1]{}

\begin{document}

\title{\system: Detection of Zero-Day Malicious Domains Leveraging Low Reputed Hosting Infrastructure}

\author{
\IEEEauthorblockN{
    Fatih Deniz\IEEEauthorrefmark{1}, 
    Mohamed Nabeel\IEEEauthorrefmark{2},
    Ting Yu\IEEEauthorrefmark{3},
    Issa Khalil\IEEEauthorrefmark{1}
} 
\IEEEauthorblockA{\IEEEauthorrefmark{1}Qatar Computing Research Institute, Doha, Qatar
} 
\IEEEauthorblockA{\IEEEauthorrefmark{2}Palo Alto Networks Inc., USA
}
\IEEEauthorblockA{\IEEEauthorrefmark{3}Mohamed bin Zayed
University of Artificial Intelligence, UAE
}
fdeniz@hbku.edu.qa, mmohamednabe@paloaltonetworks.com, ting.yu@mbzuai.ac.ae, ikhalil@hbku.edu.qa
}

\maketitle

\begin{abstract}
Internet miscreants increasingly utilize short-lived disposable domains to launch various attacks. Existing detection mechanisms are either too late to catch such malicious domains due to limited information and their short life spans or unable to catch them due to evasive techniques such as cloaking and captcha. In this work, we investigate the possibility of detecting malicious domains early in their  life cycle using a content-agnostic approach.
We observe that attackers often reuse or rotate hosting infrastructures to host multiple malicious domains due to increased utilization of automation and economies of scale. 
Thus, it gives defenders the opportunity to monitor such infrastructure to identify newly hosted malicious domains. However, such infrastructures are often shared hosting environments where benign domains are also hosted, which could result in a prohibitive number of false positives. 
Therefore, one needs innovative mechanisms to better distinguish malicious domains from benign ones even when they share hosting infrastructures. In this work, we build \system, a highly accurate practical system that not only generates daily blocklists of malicious domains but also is able to predict malicious domains on-demand. We design a network graph based on the hosting infrastructure that is accurate and generalizable over time. Consistently, our models achieve a precision 
of 99.7\%, a recall of 86.9\% with a very low false positive rate (FPR) of 0.1\% and on average detects 19K new malicious domains per day, which is over 5 times the new malicious domains flagged daily in VirusTotal. Further, \system predicts malicious domains days to weeks before they appear in popular blocklists. 
 
\end{abstract}

\renewcommand{\headrulewidth}{0.0pt}
\thispagestyle{fancy}
\lhead{}
\rhead{}
\chead{To Appear in 46th IEEE Symposium on Security and Privacy (S\&P) Conference, May 2025}
\cfoot{}

\section{Introduction}~\label{sec:intro}
Attackers increasingly utilize short-lived domains as a primary vector to launch cyber attacks, often stockpiling these domains~\cite{stockpiled} to maximize their returns. 
IBM's Data Breach Report shows that the average cost of a data breach, frequently originating from malicious domains, reached USD 4.45M in 2023~\cite{ibmcost:2022} despite numerous defense-in-depth solutions deployed to detect such malicious domains. 
Attackers keep evolving and finding more sophisticated and evasive ways to circumvent the security controls in place and lure users to access malicious websites. 
Though a plethora of detection solutions have been proposed and deployed in practice, many malicious domains either go undetected or get detected only after users are compromised. 
Thus, adapting security solutions to defend against evolving attacks is important. In this work, we leverage the fact that even though there are many possible ways to create malicious domains, the deployment options for domains are limited. 

\begin{figure}[t]
\centering
\includegraphics[width=0.95\columnwidth]{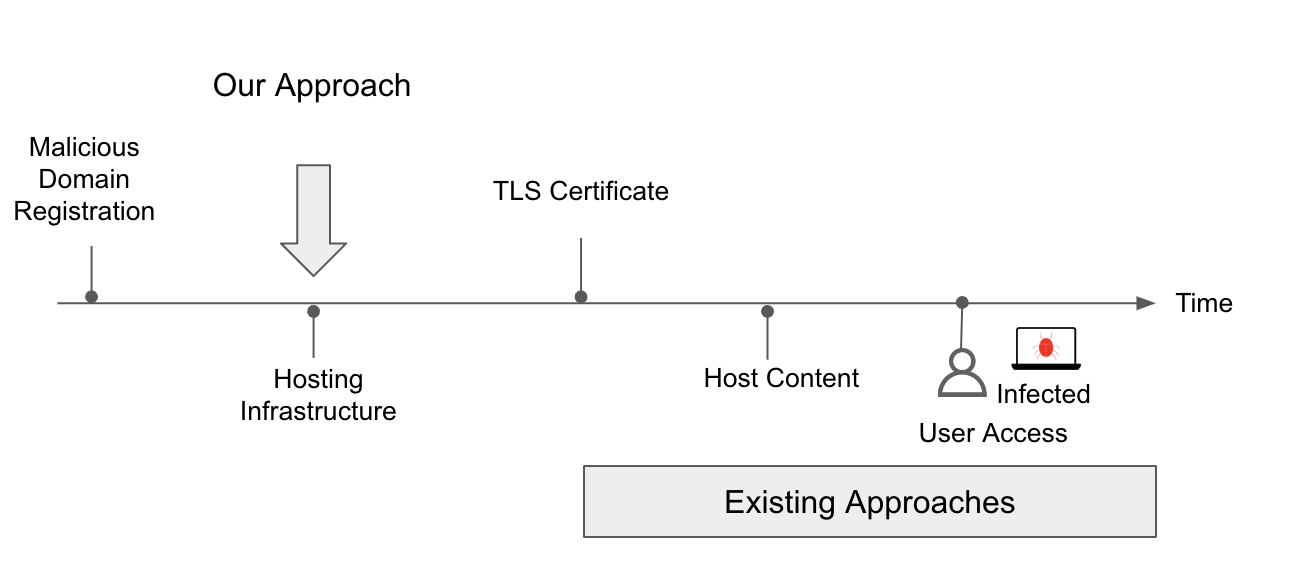}
\caption{\system vs. Existing Approaches: \system detects malicious domains much early at the hosting time compared to many of the existing techniques which often detect domains only after the web content is available.}
\label{fig:earlydet}
\end{figure}

\ccsupdate{As shown in Fig.~\ref{fig:earlydet}, the goal of our system, \system, is to detect malicious domains much early at their hosting time.
Registration time approaches result in a higher likelihood of false positives due to insufficient evidence~\cite{premadoma:euroic:2019, dga2019, vissers:euric:2019}, and most existing security scanners rely on webpage contents or network logs to detect malicious domains after they reach a security enforcement point (e.g. Firewall, browser)~\cite{contentbasedphish2021, wardman2011:contentbased,jain2020phishskape::contentbased}.
While content/network log based techniques are important, they have several limitations: (1) they have a blind spot for cloaked webpages, which is a technique attackers increasingly utilize~\cite{crawlphish:2021}; (2) they require huge amounts of computational resources to analyze billions of webpage contents; (3) there is limited visibility to malicious domains in the wild and (4) by the time malicious webpage contents or network traces are available, it is difficult, if not impossible, to prevent the attack from happening. Note that \system~\footnote{The source code and the daily generated blocklists are available at https://github.com/fatihdeniz/mantis.} is designed to augment, not replace, content-based detection methods. In some instances, like detecting compromised domains, content-based detection may be the most viable approach. 
A challenge is that we need to differentiate malicious domains from benign ones with much less available information than content-based approaches. }

\begin{figure}[t]
\centering
\includegraphics[width=0.8\columnwidth]{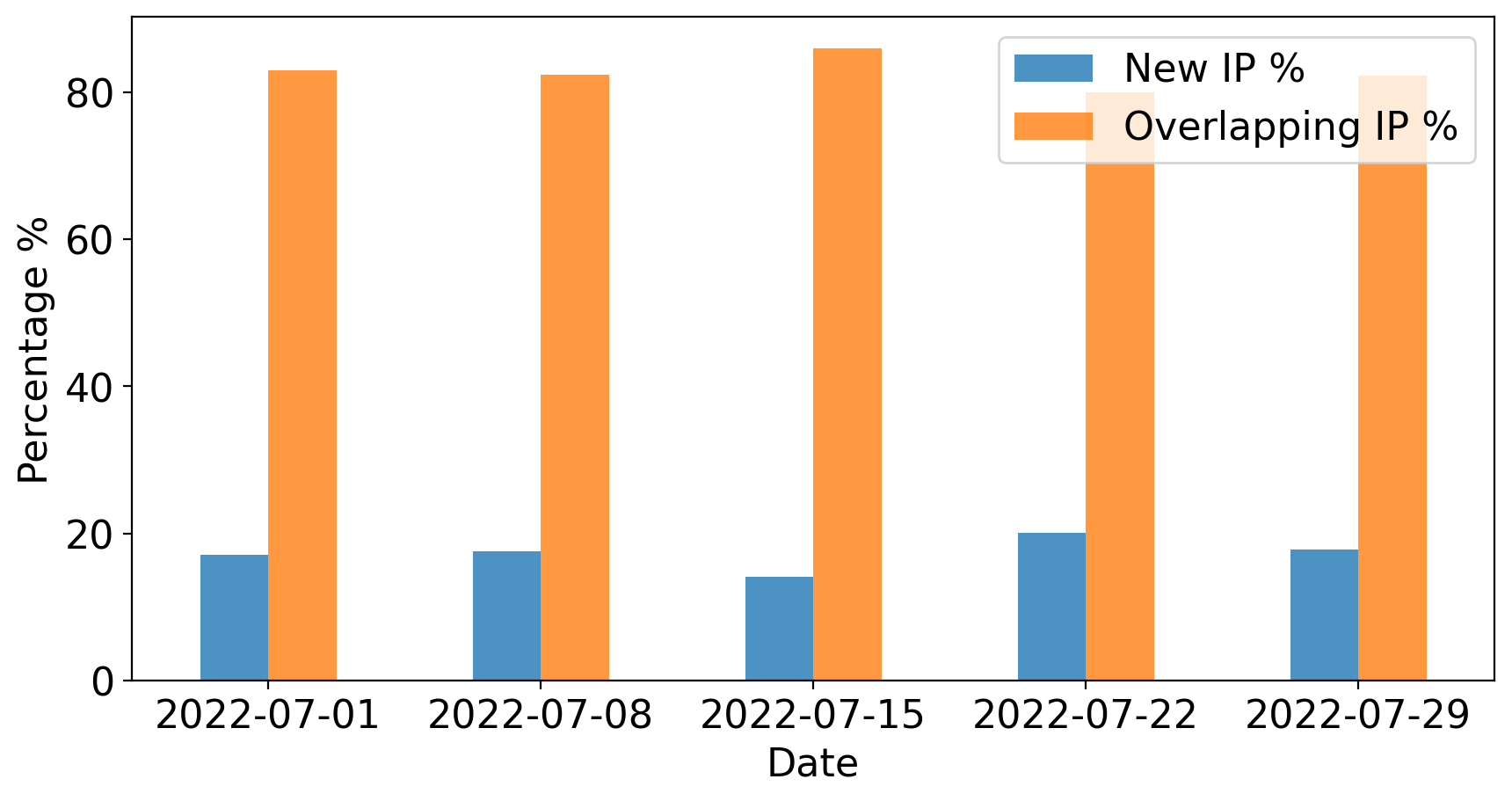}
\caption{\fatih{Reuse of Hosting Infrastructure. Over 80\% of IP addresses used to host malicious domains on a given day were found to be reused from the previous 7 days.}}
\label{fig:overlapping_ip_july}
\vspace{-2mm}
\end{figure}

We observe that while the toxicity~\revision{\cite{toxicity2013}}, i.e. the ratio of malicious domains to all domains, of hosting infrastructures on the Internet, in general, is very low, the same measure in the neighborhoods that previously hosted malicious domains is relatively high, i.e., they tend to host malicious domains again in the near future. For example, the toxicity of a sample of domains observed from passive DNS is 0.002, whereas the toxicity of a sample of domains around the IPs previously hosting malicious domains on the same day is 0.063 (31.5 times higher, detailed in Appendix~\ref{app:graphtoxicity}). 
\fatihh{Further, to evade detection, attackers deploy malicious domains with dynamic behavior by frequently rotating their IP resolutions or creating new domains. While doing so, attackers tend to reuse infrastructure resources and increasingly employ automation to host malicious domains within a similar pool of IPs (E.g. Postal campaign~\cite{campaign1:postal} and ApateWeb Campaign~\cite{campaign2:apateweb}). As shown in Fig.~\ref{fig:overlapping_ip_july}, over 80\% of IP addresses used to host malicious domains on a given day were found to be reused from the previous 7 days.}
By monitoring new domains hosted on the IPs that recently hosted malicious domains, it seems intuitive that one can identify new malicious domains easily. However, due to the increasing utilization of shared hosting, we observe that not all new domains hosted in a toxic infrastructure are malicious. In other words, being hosted on a malicious infrastructure is not conclusive evidence of the maliciousness of a domain. Therefore, additional innovative mechanisms are required to distinguish true malicious domains from false positives sharing the same hosting infrastructures.

A set of solutions extract representative features of domains and train a binary classifier to distinguish malicious and benign domains~\cite{Notos_Antonakakis2010, bilge:2014:Exposure,  lexical2015, page:2019:mal, schuppen:fanci:usenix:2018}. Yet such approaches are sub-optimal as they fail to leverage the network topology of the hosting infrastructure. Another line of solutions model domains as a graph (e.g. domain-IP graph) and exploit the label similarity of domains in the proximity of the hosting infrastructure by utilizing techniques such as belief propagation or label propagation~\cite{polonium:SIAM:2011, nazca:NDSS:2014, bp_mal2:2020, marmite:CCS:2017}. While such approaches consider network topology, they fail to capture domain features. A similar drawback exists with shallow node encoding (e.g. Node2Vec~\cite{node2vec:sigkdd:2016}) followed by a supervised classifier. 
\spupdate{It is tempting to create an ensemble classifier that combines a feature classifier with a network topology-based classifier~\cite{practicalattacks:SP:2024}. Yet, it still remains sub-optimal as it does not effectively learn the correlation between the network topology and node features.}
Graph Neural Networks (GNNs) address these issues by learning a model simultaneously considering both aspects. 
It presents a significant challenge to construct a network graph that consistently achieves high precision and recall in GNN-based malicious domain classification tasks across different time frames. Recent research suggests that belief propagation on a similar network graph outperforms GNNs~\cite{BPPhishingCCS:2022}. However, we attribute the contradictory lower performance to the deficiencies in the constructed graph, specifically the lack of preservation of the homophily property. 

A malicious domain could be compromised (i.e. benign domains exploited by attackers) or attacker-created (i.e. the domain registered by attackers). Through an analysis of daily blocklists, we found that around $82\%$ of the domains were maliciously registered, with the remaining $18\%$ identified as compromised. Since compromised domains are originally benign, they tend to be hosted on infrastructures where many other benign domains are hosted. Prior works do not distinguish between these two types of malicious domains, leading to suboptimal classification results~\cite{BPPhishingCCS:2022, ringer:ICCS:2020,hgdom:NOMS:2020}. 
Instead of building one model to detect different types of malicious domains, it is more effective to employ orthogonal detectors to identify malicious web hosting and compromised domains, such as those proposed in \cite{comp_or_at:2021:usenix}. In this study, our main objective is to detect attacker-created domains as soon as they are being hosted, meaning a DNS server resolves each domain to an IP address(es), even before it serves any content or is weaponized.

We observe that many research efforts utilizing graph models to classify malicious domains suffer from the following limitations: (1) they use biased ground truth, especially benign ground truth, where often only popular domains are utilized~\cite{Bahnsen:2017:PhishingURLsNN, practicalattacks:SP:2024}, leading to overfitted classifiers; (2) they use fully labeled graphs and are not shown to work on unlabeled domains under a semi-supervised learning setting; 
and (3) they are unable to perform on-demand predictions on domains that are not part of the training graph~\cite{BPPhishingCCS:2022, bp_mal2:2020}. 
We address these issues and build a practical system named \system
that support two key use cases: (1) daily blocklist generation (batch mode); and (2) on-demand prediction. On a daily basis, \system first compiles a list of seed malicious domains first seen on a given day and identifies other recent domains hosted on the same infrastructure as the seed domains.
Based on these resolutions, \system builds a graph and then collects lexical and hosting features and ground truth domains to train a model. 
An ensemble of daily trained models is utilized to predict malicious domains not present in the training graph to further reduce false positives. \ccsupdate{To the best of our knowledge, we are the first to construct a graph around \emph{attacker-owned} domains that yields consistent results over different and temporarily separate datasets.}

\ccsupdate{
In summary, we make the following contributions.
We devise a method to automate the construction of a graph around attack domains, enabling the identification of new ones. The guided graph expansion facilitates the creation of high-toxicity graphs in the ever-increasing utilization of shared hosting and reduces graph size significantly (around $5\%$ of the all-active-resolutions graph). \spupdate{With this targeted perspective, we detect malicious domains with a low FPR days to weeks ahead of popular detectors.} We construct representative benign and malicious ground truth, leading to accurately predicting the maliciousness of domains not seen in the input graph, which is often overlooked in domain classification. We apply inductive training that 
enables on-demand predictions, even when no labeled domain exists within the computation graph.
To enhance transparency and interoperability, we provide feature importance, shedding light on the significance of different feature categories and analyze our robustness. We build a sanity checking system to monitor the predicted malicious domains and show that \system serves as an early detection system. 
As a testament to its practicality and effectiveness, \system has been operational for over a year and detects $\sim$19K malicious domains (5 times more than VirusTotal malicious domains) daily at a low FPR of 0.1\%. 
}

\section{Background and Data Sources}~\label{sec:background}
\vspace{-5mm}
\subsection{Graph Neural Networks (GNNs)} \label{ss:GNNs}

GNNs are a class of deep learning models for learning from data represented as graphs. GNNs learn representations of either nodes, edges, or whole graphs. In this work, as our objective is to detect malicious domains, we focus on node representation learning. 
\fatih{GNNs combine node features with the graph
structure by recursively passing neural messages along the
edges of the input graph using three key functions, namely \texttt{MSG}, \texttt{AGG}, and \texttt{UPDATE}. These functions work together for exchanging messages between a node $v_i$ and its immediate neighboring nodes $\mathcal{N}_{v_i}$. 
In layer $l$, a message between two nodes $(v_i, v_j)$ depends on the previous layer's hidden representations $h_i^{l-1}$ and $h_j^{l-1}$, i.e, $m_{ij}^l = \texttt{MSG}(h_i^{l-1}, h_j^{l-1})$. 
\texttt{AGG} combines the messages from $\mathcal{N}_{v_i}$ with $h_i^{l-1}$ to produce $v_i$'s representation for layer $l$ in  \texttt{UPDATE}. 
Various adaptations of this core message passing framework with different \texttt{MSG}, \texttt{AGG}, and \texttt{UPDATE} implementations have been proposed~\cite{graphsage:nips:2017, hgt:web:2020, gcn:iclr:2017, rgcn:www:2018, gat:velickovic2018graph}. These can be categorized into two groups: those that work on homogeneous graphs~\cite{graphsage:nips:2017, gcn:iclr:2017, gat:velickovic2018graph} with one type of nodes and edges and those on heterogeneous graphs~\cite{hgt:web:2020, rgcn:www:2018} with different node and/or edge types. To harness the comprehensive representation, we employ a heterogeneous GNN that considers both node and edge types.}

\subsection{Data Sources}
\eat{In this section, we provide a brief description of the data and intelligence sources we use in this work.} We utilize the following data sources:

\textbf{Passive DNS:}  
\revision{We use Farsight PDNS data~\cite{DNSDB} that is collected from sensors placed near DNS resolvers and provides a summary of domain resolutions and publicly accessible updates to zone files. Farsight streams records at various time granularity, ranging from every minute to daily to monthly. For training purposes, we use the daily feed, but for on-demand detection, we advise using the feed published every minute.} \fatih{We use PDNS to extract domains related to seed domains (expansion) and collect domain/IP features.} \eat{In our research, we use Farsight data as our PDNS feed. One advantage of PDNS is that it preserves the privacy of individual Internet users as it contains only aggregated information. However, such data is not as rich in information as proxy/HTTP DNS logs, which not only contain individual DNS queries and responses but also timing information. Despite its limited information, we are able to extract domains related to seed malicious domains as well as domain/IP features from the PDNS repository as described in Section~\ref{sec:approach}.} \eat{Previous research utilizes this data feed to uncover the behavior of domains in the wild and also detect malicious domains~\cite{Kintis:2017:Combosquatting, Khalil:2018:Inference}.} \eat{In this work, we use the following three record types from PDNS:
\begin{itemize}
    \item SOA records: they contain the MNAME (the primary name server for the domain) and RNAME (email of the domain name administrator). 
    \item A records: these records map the domain names to their IPs. 
    \item MX (mail exchange server) and NS (name server) records: these records specify where inbound mails for a domain should get directed and authoritative name servers respectively.
\end{itemize}

All of the above records contain aggregated timing information (such as the first and last seen timestamps for each record), and the count of DNS lookup requests. Such timing information is crucial to identify recent records to build our graphs and perform classification.}

\textbf{VirusTotal URL Feed (VT):} ~\cite{VirusTotal}
\eat{URL feeds have become essential in the evolving security technology landscape. Organizations integrate their Security Orchestration, Automation, and Response (SOAR) products with threat intelligence platforms like VirusTotal to gather insights about encountered URLs. This integration leverages contributions from a vast network of global sensors connected to email and internet gateways, enabling real-time visibility into potential and emerging threats.}
VT feed contains all the URLs queried by users worldwide and VT provides an API to monitor the status of specific URLs and generates an hourly feed of JSON-encoded reports. Using our organization's subscription, we observe a daily volume of approximately 2-5 million URLs within VT's feed in Sep 2023, monitoring them regardless of their maliciousness.
This feed comprises URLs and aggregated intelligence collected from over 90 third-party scanners, including notable sources such as Google Safe Browsing (GSB)\footnote{GSB is available for enterprises through cloud-based Google WebRisk}~\cite{GoogleSafeBrowsing}, Phishtank~\cite{phishtank}, among others. 
We continuously profile newly observed domains in this feed and utilize them in two primary ways: (1) to generate a daily collection of potentially malicious seed domains, and (2) to construct our malicious ground truth. Following the prior studies~\cite{phicious:raid:2022,comp_or_at:2021:usenix}, we set a threshold of five scanners to identify malicious domains, which allows us to aggregate intelligence from diverse scanners, enhancing the generalizability of our dataset. \eat{If a domain is reported as malicious by five or more scanners and determined to be an attack domain, it is included in our ground truth. This approach allows us to aggregate intelligence from diverse scanners, enhancing the generalizability of our dataset.}

\revision{\textbf{Alexa, Tranco, Cisco Umbrella, and Google’s Chrome User Experience Report (CrUX) Top Lists:}   \fatih{To ensure a comprehensive and unbiased benign ground truth, we employ a robust approach that integrates multiple reputable top lists along with randomly selected benign domains from passive DNS based on predefined rules. By leveraging diverse sources like Alexa~\cite{Alexa}, Umbrella~\cite{Umbrella}, Tranco~\cite{trancolist}, CrUX~\cite{Crux}, and passive DNS, we aim to minimize any inherent bias and provide a more accurate representation of benign domains.}
Alexa lists the most popular 1 million domains each day. 
Umbrella aggregates DNS queries to the OpenDNS resolvers to create the most frequently queried names.
Tranco~\footnote{Available at https://tranco-list.eu/list/Y52LG.} aggregates rankings from the Alexa, Umbrella, and Majestic~\cite{Majestic} lists, and CrUX computes rankings directly based on browsing data from Chrome users and provides ranking buckets (e.g., top 1K, 5K, 10K, etc.). 
\eat{It is aggregated monthly, and as suggested by~\cite{toppling2022}, it captures the unordered set of most popular websites more accurately than other top lists.} }
Indeed, a popular domain does not always mean it is benign~\cite{toppling2022}.
However, domains consistently appearing in the top lists over a period are highly likely to be benign, as attackers tend to use a domain for a short time period, their domain popularity is likely to last only a few days~\cite{bilge:2014:Exposure, toplistsanalyze:2019}.
Based on these observations, following prior studies, for the daily top lists from Alexa and Umbrella, we compile the top 30-day list, which includes the domains consistently appearing in the list for 30-days and also exclude those marked as malicious by VT. They are combined with the monthly lists from Tranco and CrUX to form part of our benign ground truth.

\section{Approach Overview}~\label{sec:overview}
\vspace{-5mm}

\subsection{Daily Blocklist Generation}
\label{subsec:dailyblocklistgeneration}

\begin{figure}
\centering
\includegraphics[width=1\columnwidth]{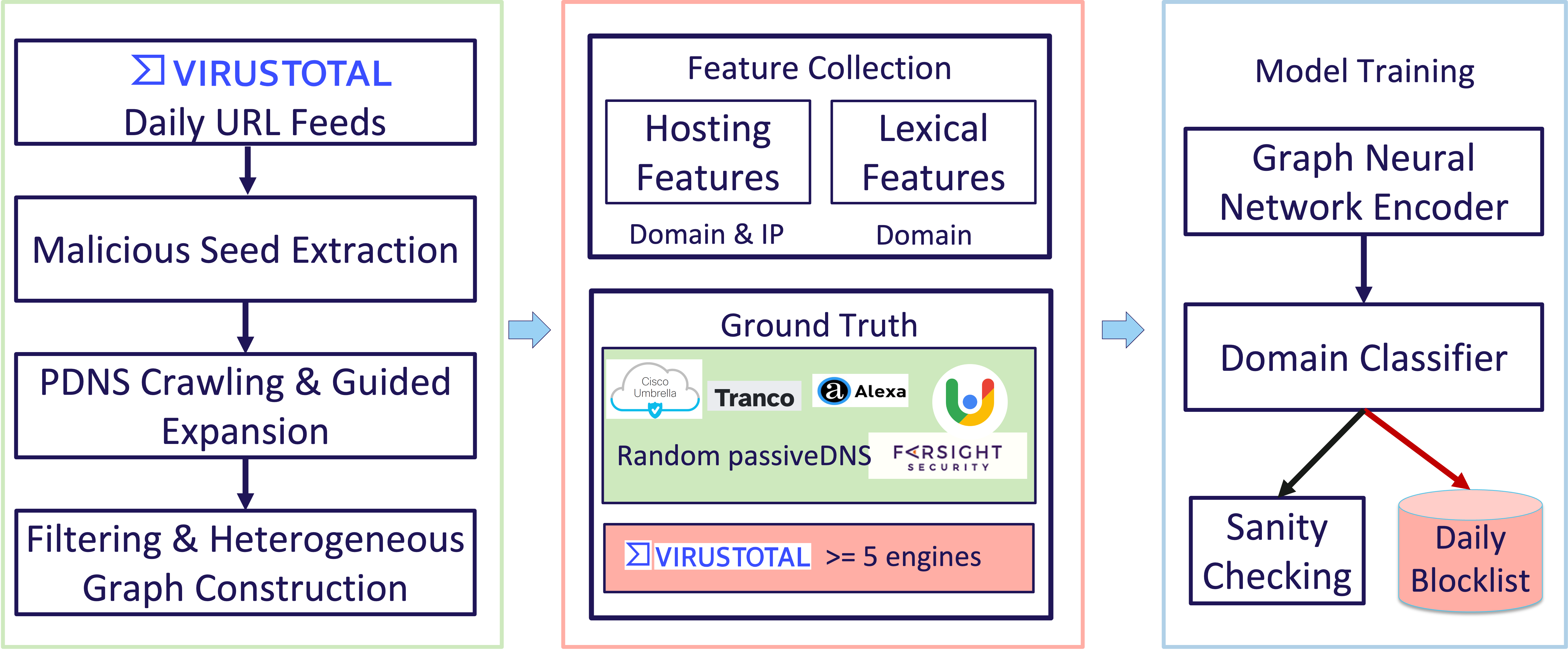}
\caption{Overall pipeline for daily blocklist generation.}
\label{fig:pipeline}
\vspace{-2mm}
\end{figure}

Fig.~\ref{fig:pipeline} shows the overall pipeline of detecting malicious domains for a given day. 
On a given day, our initial step involves identification of a collection of attacker-owned malicious seed domains that were first seen that day. This process relies on analyzing the daily blocklisted URLs in conjunction with our internal VT domain dataset. This step is critical in our capacity to identify and monitor potential malicious activities within the dynamically changing threat landscape.
Then, following the observation that attackers often reuse hosting infrastructure to launch their attacks, we crawl PDNS of recently hosted domains in the hosting neighborhood of seed malicious domains to construct a graph based on PDNS records. 
By automating the construction of a graph centered around attack domains (excluding compromised domains) and guiding its expansion, we generate a highly toxic graph. In our guided expansion, instead of utilizing all resolutions, we identify only the most recent resolutions to expand the graph. Further, we rank the resolutions by resolution time and select the most recent $N$ resolutions. This guided expansion not only assists in building a graph with high toxicity in the ever-increasing utilization of shared hosting but also reduces the size of the graph significantly (around 5\% of the size of all active resolutions). 
Notably, $16.65\%$ of domains within the unknown segment of our graphs receive a malicious classification from at least one VT engine, underscoring our ability to identify highly malicious domains. The distribution of VT positives in a randomly sampled expanded graph is provided in Appendix~\ref{app:graphtoxicity}.
Our heterogeneous graph consists of apex domains (i.e., e2LDs), fully-qualified domain names (FQDNs), IPs, subnets~\footnote{Our empirical analysis shows that that subnet 24 yields the most favorable classification results, and thus we set the subnet size to 24} and Autonomous System Numbers (ASNs). As to node features, besides lexical features of domain names, we also collect a set of novel hosting features for both domains and IPs. When PDNS lacks information on certain domains, it leads to the absence of \textit{all} related features. Since existing imputation techniques assume the availability of some features and thus do not work~\cite{grape2020}, we propose our own solution. 

For malicious ground truth, we use a subset of the malicious seed nodes as well as previously seen malicious domains within the expanded graph. For benign ground truth, we take a pragmatic approach and compile a representative ground truth by considering multiple sources. Most prior research utilizes only Alexa or Umbrella top list as the benign ground truth~\cite{Bahnsen:2017:PhishingURLsNN, bp_mal2:2020}, which represents a biased set of benign domains and inevitably results in models with high false positives in practice due to several reasons, such as the exclusion of benign domains with low web traffic.  With the constructed heterogeneous graph and the ground truth, we train a semi-supervised GNN to predict unseen malicious domains in the neighborhoods of seed malicious domains. 

\subsection{On-Demand Prediction}
\begin{figure}
\centering
\includegraphics[width=0.90\columnwidth]{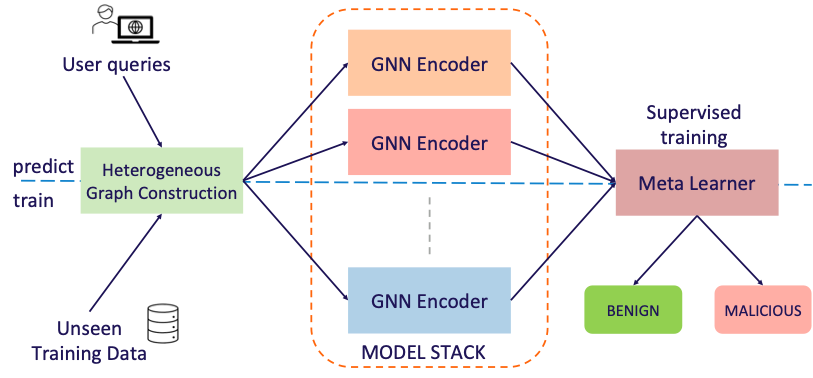}
\caption{On-demand detection of malicious domains.}
\label{fig:realtime}
\vspace{-2mm}
\end{figure}

A key deficiency in existing graph-based malicious domain detection solutions~\cite{bp_mal2:2020, hgdom:NOMS:2020} is that they cannot predict the maliciousness of a domain that does not appear in the training graph.
Since re-training a graph model is computationally expensive and is not practical,
an inductive approach, which trains a model on one graph and then can apply the model to a different graph without re-training, is much desired in a practical system.
For a new domain, we construct the passive DNS graph around the neighborhood of this domain (i.e., the target domain computational graph) and perform only a forward pass to obtain the embeddings from these stacked models. We utilize an ensemble classifier by employing a stack of semi-supervised GNN models and a meta-learner to combine the embeddings from these models for the final classification.
Fig.~\ref{fig:realtime} illustrates how we perform on-demand detection of domains not in our domain graph in an inductive manner.

\section{Detailed Design of \system}~\label{sec:approach}

\subsection{Ground Truth Collection}~\label{sec:gt}\textbf{Seed Extraction:}
We collect two sets of malicious domains through VT. The first is a set of seed malicious domains, which are used to construct a network graph. The other is the malicious domain ground truth for model training. 
For the daily seed domains, we first select those URLs seen for the first time in the VT feed within the past 24 hours and marked as malicious by at least 5 scanners and an active VT scan of the corresponding domains with at least 3 VT scanners following prior research~\cite{comp_or_at:2021:usenix}. 
\fatihh{We extract only those malicious domains that are highly likely to be created by attackers and filter out potentially compromised domains (whose apex domains appear in the top lists utilized in this work) as well as domains belonging to web hosting services (e.g., weebly.com). We employ a rule-based approach to differentiate between public domains (such as web hosting and CDN domains) and attacker-created domains~\footnote{We use the whitelists from Mozilla’s widely accepted Public Suffix List~\cite{publicsuffix:mozilla} and the data from~\cite{comp_or_at:2021:usenix} available at \url{https://bit.ly/3Yw8Sip}.}, which is widely used in the industry to flag malicious domains~\cite{paloalto:2014}.} \eat{Further, we exclude sinkhole domains as their hosting infrastructure differs from attack infrastructure.
We further use an LSTM-based model to filter out DGA domains as they are not the focus of our detection and there exist highly effective techniques to detect DGA domains.} 
From the remaining domains, we identify those with popular brand keywords~\cite{phicious:raid:2022} and include them within the seed domains. At the end of this process, we identify \revision{close} to 3000 likely attack seed domains per day seen for the first time from June 1st, 2022 until June 1st, 2023.

\fatihh{\textbf{Malicious GT:} We utilize a subset of seed domains ($VT \ge 5 $) in our ground truth. Also, after constructing the network graph from these seeds, we identify previously marked malicious domains using the same threshold within the expanded graph and incorporate them into our malicious ground truth.}

\eat{MOVED TO APPENDIX: Once we construct the network graph from the seed domains, we randomly select a set of domains from the graph and follow the below mentioned manual process to identify those malicious ones as our malicious domain ground truth. 
\fatih{During manual ground truth collection, we investigate the presence of phishing and fraudulent activities, distribution of malware, malicious or harmful content, and involvement in different types of brand squatting attacks. In addition to investigating the website content, we assess auxiliary information such as registration information, hosting information, Internet Wayback Machine snapshots, and historical WHOIS records. We also assess the detailed reports from two threat intelligence platforms, Microsoft Defender
Threat Intelligence (ti.defender.microsoft.com) and AlienVault (otx.alienvault.com).}}

\textbf{Benign GT:}
\spfinal{
Benign domains are randomly collected from the domains in the expanded PDNS graph, and VT is used to further ascertain their benignness. We leverage diverse sources, including top lists, existing benign sources, and reputable TLDs. To reduce bias, we focus on newly seen domains in the VT feed and unpopular domains, as Tranco and other domain ranking lists tend to skew toward popular, long-established domains. Specifically, we randomly sample 10,000 domains from the expanded graph that do not exist in our local VT domain database or exist in the local unpopular domain dataset, such as those collected from sources like Yellow Pages~\cite{YellowPages}.
We perform additional heuristic-based filtering to remove suspicious domains, including those resolved to sinkhole IPs, those with invalid or expired certificates, or lacking valid content.
We also exclude DGA domains, domains impersonating popular brand names, and those from TLDs managed by Freenom\footnote{Freenom was discontinued in 2023, and therefore this filtering step no longer removes a significant number of domains.}—such as .gq, .ml, .cf, and .tk—as they generally have very low reputations. We then submit the remaining domains to VT to check their status, and extract those marked as benign by all scanners. While these heuristics do not offer a precise or comprehensive method for detecting malicious domains with high confidence, they help build a high-quality benign ground truth to ensure robustness in classification.
In addition to domains identified through active lookup, we passively collect .edu and .gov domains, as well as top lists from sources like Alexa, Umbrella, CrUX, and Tranco. Since relying solely on top lists leads to poor generalization and a higher false positive rate, we integrate newly observed domains, domains from reputed TLDs, and other top lists into the benign ground truth to create a more representative dataset, enabling us to train a generalizable classifier with a lower false positive rate. When we examine the distribution of our benign ground truth across different categories, we observe that approximately 54\% of the benign ground truth comes from popularity-based domains, 41\% from heuristics-based domains, and 5\% from educational and governmental websites.
Further details related to the quality checks on the ground truth are provided in Appendix~\ref{app:sanity_table}. 
}

\eat{\subsection{Hosting Infrastructure Expansion}~\label{sec:datacollection}
As described above \repe{ in Section~\ref{sec:gt}}, we construct the seed domain list from the VT URL feed. Then, using the passive  DNS database, we expand the seed domain list by identifying their hosting IPs, then other domains recently hosted on those IPs and finally the hosting IPs of these new domains. It is important to build an appropriate-sized graph in terms of classification performance (F1-score) and computational cost in order to run the model in practice. As evaluated in Section~\ref{subsec:pdnsexpansion}, we empirically identify that the best number of hops is 3 from the seed domains and the best number of recently hosted domains on each IP is 200.}

\subsection{Graph Construction}
~\label{subsec:graphconstruction}

\begin{figure}[t]
\centering
\includegraphics[width=0.29
\textwidth]{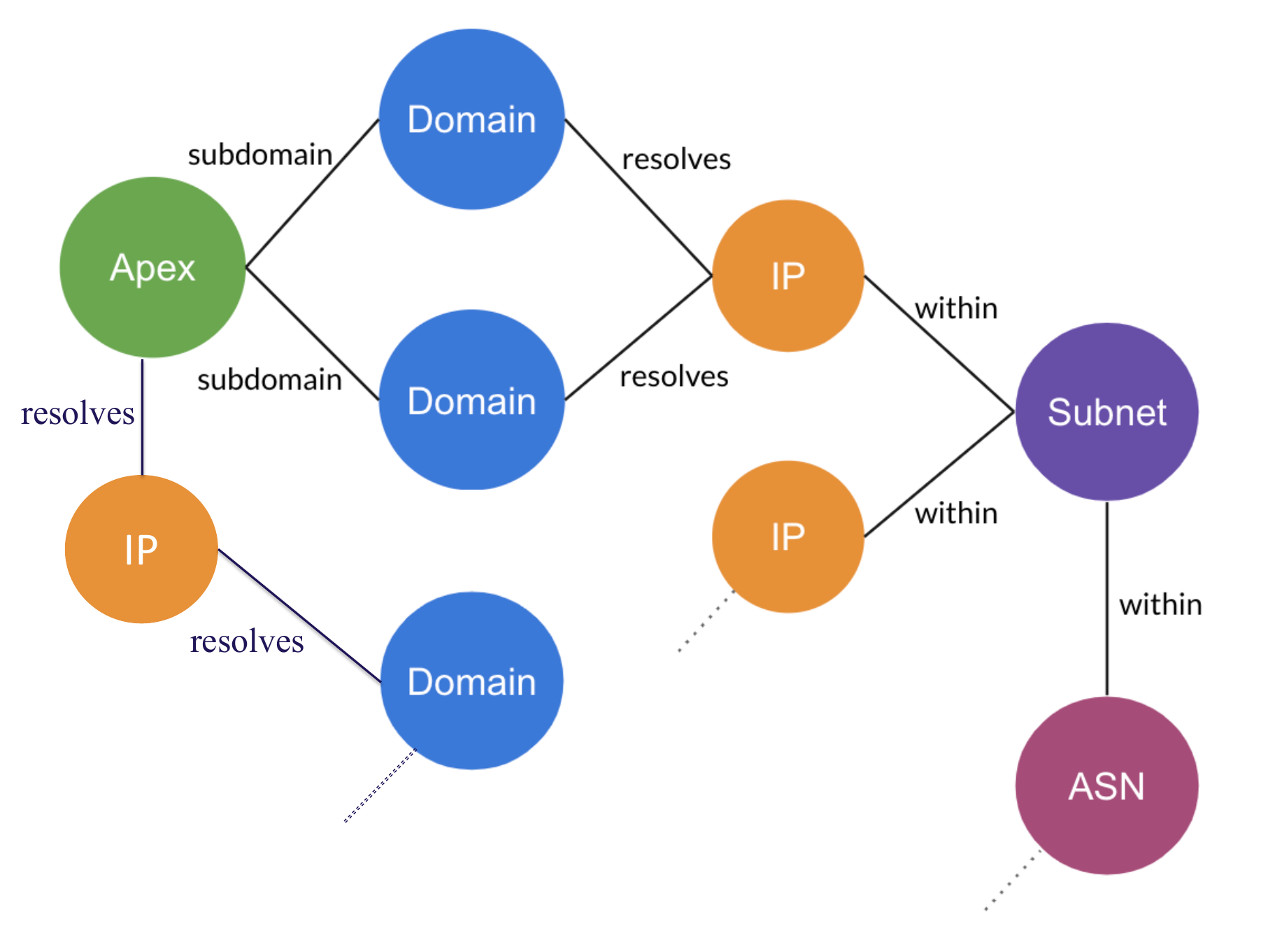}
\caption{Graph schema.}
\label{fig:schema}
\end{figure}

As described above \repe{ in Section~\ref{sec:gt}}, we construct the seed domain list from the VT URL feed. Using the passive DNS database, we expand the seed domain list by identifying their hosting IPs, then other domains recently hosted on those IPs, and finally, the hosting IPs of these new domains. Fig.~\ref{fig:schema} shows the graph schema we construct using the DNS resolutions. This heterogeneous graph consists of domains (both apexes and FQDNs), IPs, subnets, and ASNs. Domains are connected based on subdomain relationships, and they are linked to IP nodes to which they resolve.
IP nodes are connected to Class C subnets, which are further connected to their respective ASNs. We perform the following graph pruning to 
focus more on attacker owned infrastructure. While such pruning improves the classification performance of the model, the reduction of the number of nodes due to pruning is quite low ($<1\%$).
In order not to negatively impact the generalizability and effectiveness in practice we do not use any filters like Segugio~\cite{Segugio_Rahbarinia2015} that filters domains with low numbers of queries, and Exposure~\cite{Exposure_Bilge2011} that removes domains that are older than one year.

\begin{itemize}[leftmargin=*]
\itemsep0em 
    \item To ensure an accurate depiction of the attack infrastructure, it is crucial to identify the attack neighborhood at the time of the attack. This is particularly important as attack domains are often sinkholed after the attacks. In our approach, we leverage passive DNS to check the domain resolutions and exclude sinkhole IPs~\footnote{We utilize the SinkDB~\cite{sinkDB:2024} and MISP Project~\cite{MISPSINKHOLE:2024} sinkhole lists.}. By doing so, we ensure that our depiction of the attack infrastructure is precise and aligns with the actual scenario.
        \item Since public domains, such as \url{blogspot.com}, reduce the strength of the homophily relationship, we prune them$^6$.
    \end{itemize}

\subsection{Feature Engineering}~\label{subsec:featureengineering}
\ccsupdate{In our pursuit of creating a practical system, we purposefully select features that are widely accessible and commonly employed. \spupdate{We ensure that attributes used in ground truth selection are excluded from our feature sets. For example, the utilization of popularity rankings in both selecting the benign ground truth (with potential filtering of unpopular domains) and within the feature sets~\cite{practicalattacks:SP:2024} may inflate the perceived effectiveness, yet proves ineffective in practice.}
Our feature set encompasses three main categories: lexical, domain hosting, and IP features.} 
The lexical features focus on the domain name itself, capturing attributes such as suspicious keywords, length, and specific character patterns. The domain hosting features shed light on the hosting environment, including factors like access frequency, the presence of multiple IP addresses and name servers, and the consistency between domain apex and name servers and IP features provide insights into the IP addresses associated with the domain, including number of apex domains hosted, access frequency, and the duration of appearance in PDNS records. The rationale is that short-lived, recently created domains with sporadic access patterns are more likely to be malicious. 
We enhanced the existing features by introducing novel IP features and a few lexical features. The details of these features are provided in Table~\ref{tab:features} in Appendix~\ref{app:appnodefeatures}.
Further insights and statistics into the significance of features are provided in Section~\ref{sec5:feature_importance}. 
\ccsupdate{Based on the explanations, IP features, in particular, have significant contributions to the predictions.}

An additional challenge arises when certain domain and IP nodes lack hosting features because the PDNS database has not recorded any resolutions. This could occur as we empirically use a shorter window of 7 days to extract features, and during that window, PDNS does not have any records on them. Unfortunately, existing feature imputation techniques do not apply to these missing features, as all related features are missing~\cite{grape2020}. Hence, we devise a novel feature imputation technique leveraging the graph structure, with the intuition that nodes closer to one another tend to have similar characteristics. We select the five closest neighbors from the node's neighborhood and calculate the weighted average of their features to determine the node's features with higher precision. A higher number of shared IPs suggests a stronger connection between the two domains. We apply the same rationale to identify the association between two IPs if they host several common domains.

\subsection{Malicious Domain Classifier}~\label{subsec:modeltraining}
To experimentally validate our approach, we choose three temporally separated datasets. We train the classifier on one-week data and test it on the following day's data. Table~\ref{tab:dataset} provides the statistics of these datasets.

\begin{figure}
\centering
\begin{subfigure}[t]{0.40\columnwidth}
    \includegraphics[width=\columnwidth]{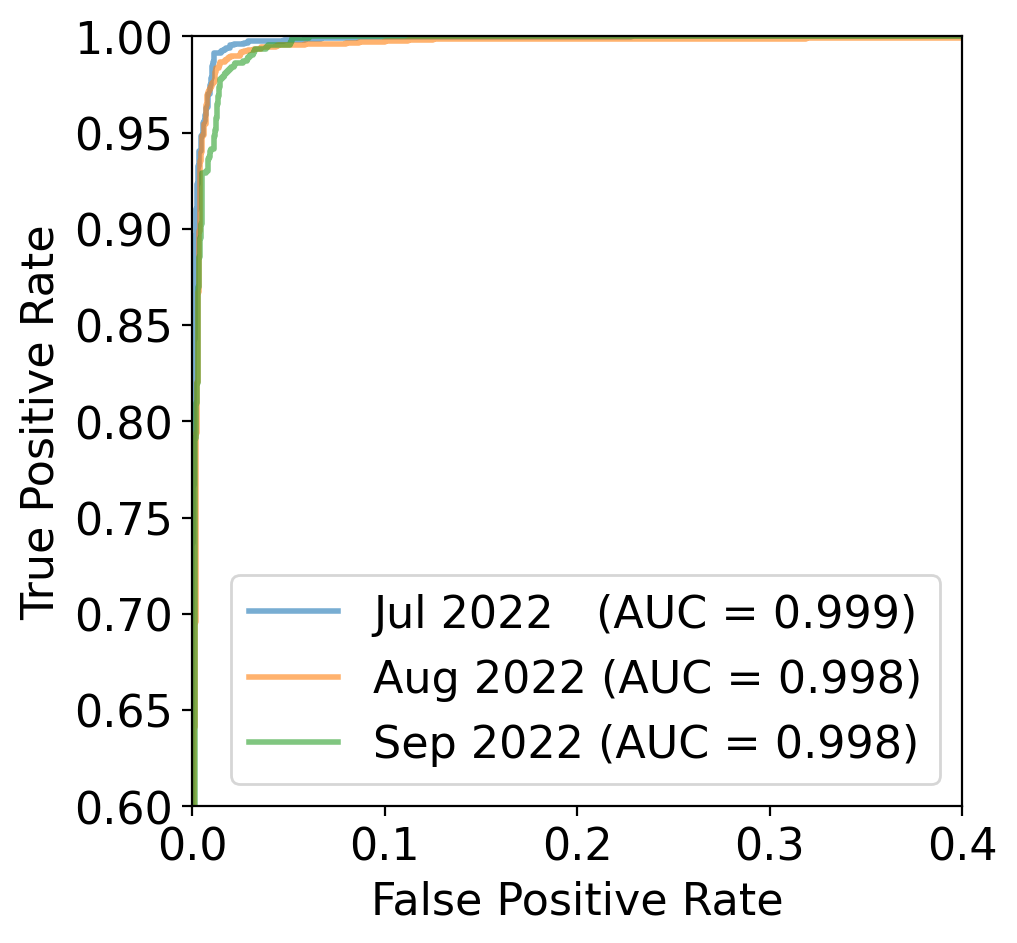}
    \caption{}
    \label{5roc:training}
  \end{subfigure}
  \begin{subfigure}[t]{0.40\columnwidth}
    \includegraphics[width=\columnwidth]{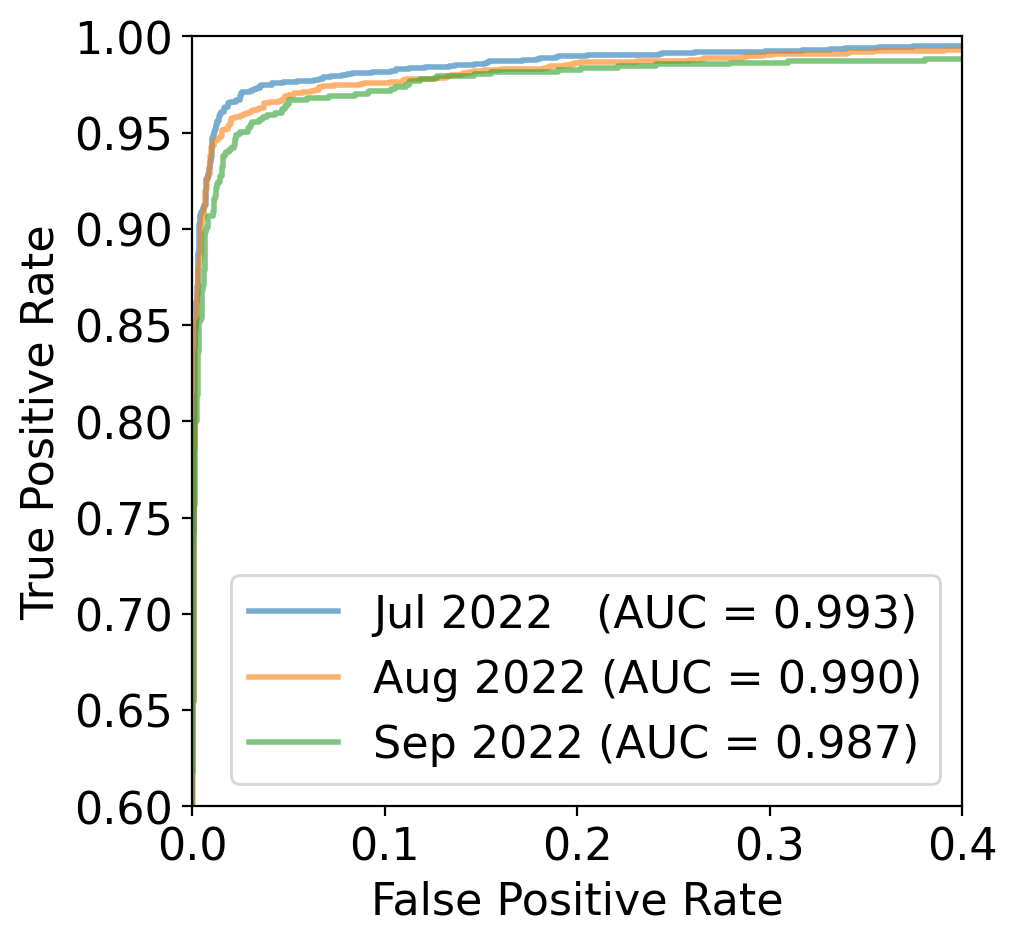}
    \caption{}
    \label{5roc:testing}
  \end{subfigure}\hfill
  \caption{(a) Validation, (b) Testing ROCs for daily blocklist.}
  \label{5fig:rocs}
\end{figure}

\textbf{Model Training:} With the constructed graph in Section~\ref{subsec:graphconstruction}, we assign the features extracted in Section~\ref{subsec:featureengineering} to each node and inject the benign and malicious labels collected in Section~\ref{sec:gt}.  Using 5-fold cross-validation, we deploy a semi-supervised multi-relational GNN that can incorporate information by taking into account node and edge types~\cite{schlichtkrull:multignn2018}. Our GNN comprises three layers with embedding dimensions of 256. We utilize a learning rate of 0.01, employ neighbor sampling, and a final layer that aggregates all embeddings from the preceding layers.
Fig.~\ref{5roc:training} displays the ROC curves, while Table~\ref{tab:modelvaltest} presents the precision, recall, and F1 scores for the three validation datasets at a low FPR (0.5\%). We observe that our approach consistently performs well over all the datasets, demonstrating the generalizability of our model in different time periods. 
\begin{table}
\centering
\caption{Datasets used in blocklist generation experiments.}

\resizebox{\linewidth}{!}{
\begin{tabular}{c|c|c|c|c}
\toprule
\textbf{Dataset} & \textbf{\#Domains} & \textbf{\#IPs} & \textbf{\#Malicious} & \textbf{\#Benign} \\ 
\midrule

Jul 01-07 2022 Train & 794728 & 110500 & 10185 & 13964 \\ 
Jul 08 2022 Test & 134160 & 45865 & 856 & 2137 \\ 
\midrule
Aug 01-07 2022 Train & 702989 & 99104 & 8122 & 12823 \\ 
Aug 08 2022 Test & 126370 & 33883 & 842 & 1659\\ 
\midrule
Sep 01-07 2022 Train &  480654 &  74549 & 4462 &  12724         \\ 
Sep 08 2022 Test & 118811 & 41481 &  766 & 1842 \\ 
\bottomrule
\end{tabular}
}
\label{tab:dataset}
\end{table}

\textbf{Model Testing:} During the testing phase, we take the labeled domains observed on the following day of the training window. For example, if the model is trained on July 01-07 2022 window, the testing data is collected from July 08, 2022. We ensure our testing domains do not appear in the training labels to avoid data leakage. We build a graph around these testing domains similar to how we build the training graph and append it to the training graph before performing the forward pass. Fig.~\ref{5roc:testing} shows the ROC curves, while Table~\ref{tab:modelvaltest} displays the testing performance results at a low FPR (0.5\%).
We see that our model consistently achieves high precision across temporally different datasets at testing time. We train this classifier daily and utilize it to proactively generate a daily blocklist of malicious domains.

\subsection{On-Demand Malicious Domain Classifier}~\label{ss:ondemand}
The purpose of the on-demand classifier is to assess the maliciousness of \emph{any} domain in the wild. Thus, its goal is different from blocklist generation. In blocklist generation, we know that our seed nodes are highly likely to be malicious and within the computation graph of each node there exists at least one malicious node. By learning the behavior of these malicious nodes, \system infers unseen malicious domains. However, relying only on one graph for an on-demand classifier could result in high false positive rates as the domain being classified may have a different distribution than those in the training dataset. One way to reduce this distribution gap is to consider training data from multiple periods so that the training data is likely to capture the distribution of domains in the wild. Thus, we design a different classifier for the on-demand classification with the following process.

\eat{\todo{We decided to use this approach after checking the predictions for the Alexa 100K 1week domains. We should maybe mention what we faced there, or keep it abstract.} \nabeel{Fatih, Let's add this detail here - these practical steps add value and help others to replicate our work.}}

\eat{\todo{Include  some examples like: adsforcomputercity.com, adslivetraining.com, notiftravel.com https://redecanais.cx/ https://winluckychance.com	from alexa 100k30d}}

\eat{\todo{Let's use one section to summarize the on-demand meta-learner. Describe the following here:}}

\textbf{Data Collection:} For the on-demand classifier, we utilize semi-supervised inductive models trained for blocklist generation. Instead of using the resulting confidence scores, we collect the embeddings from the last GNN layers and train a meta-learner with these embeddings. Thus, the data for the on-demand classifier is the same as for the blocklist classifier, except that a longer training window is considered.

\textbf{Ground Truth Collection:}
The on-demand classifier utilizes the malicious and benign ground truth from daily collections. It combines the validation sets of the weekly trained models with the following months' ground truth, which is not part of the training graphs. For example,  we use the weekly models for July and select the benign and malicious seeds from the following month (August) that are not part of these training graphs as our ground truth data for the meta-learner. In this way, we expect on-demand classifiers to perform and adapt well to the rapidly evolving environment.

\begin{table}
\centering
\caption{Daily blocklist performances \review{at 0.5\% FPR thresh.} \eat{\nabeel{Fatih: if you have the classification results, can we update this table to show precision and recall at 0.1\% FPR? If we do that, we don't need the FPR column.}}}
\footnotesize
\begin{tabular}{l|l|c|c|c}
\toprule
&\textbf{Dataset} & \textbf{F1} & \textbf{Precision} & \textbf{Recall} \\ 
\midrule
\multirow{3}*{Validation}
    &Jul 01-07 2022 & 0.975 & 0.990 & 0.960 \\ 
 &Aug 01-07 2022 & 0.972 & 0.991 & 0.954 \\ 
 &Sep 01-07 2022 & 0.954 & 0.981 & 0.931 \\ 
\midrule
\multirow{3}*{Testing}
    &Jul 08 2022 & 0.954 & 0.990 & 0.921 \\ 
 &Aug 08 2022 & 0.953 & 0.989 &  0.920 \\ 
 &Sep 08 2022 & 0.938 & 0.981 & 0.898 \\ 
\bottomrule
\end{tabular}

\label{tab:modelvaltest}
\end{table}

\begin{figure}
\centering
\begin{subfigure}[t]{0.40\columnwidth}
    \includegraphics[width=\columnwidth]{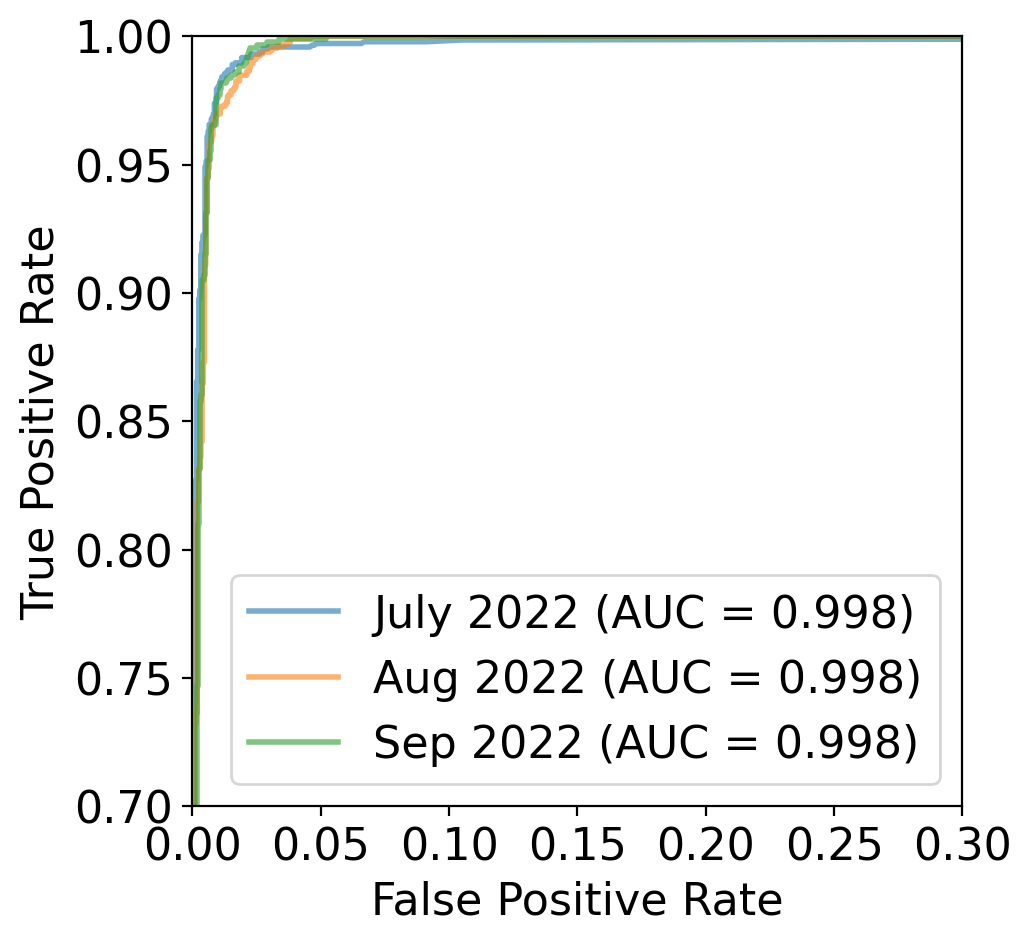}
    \caption{}
    \label{6roc:training}
  \end{subfigure}
  \begin{subfigure}[t]{0.40\columnwidth}
    \includegraphics[width=\columnwidth]{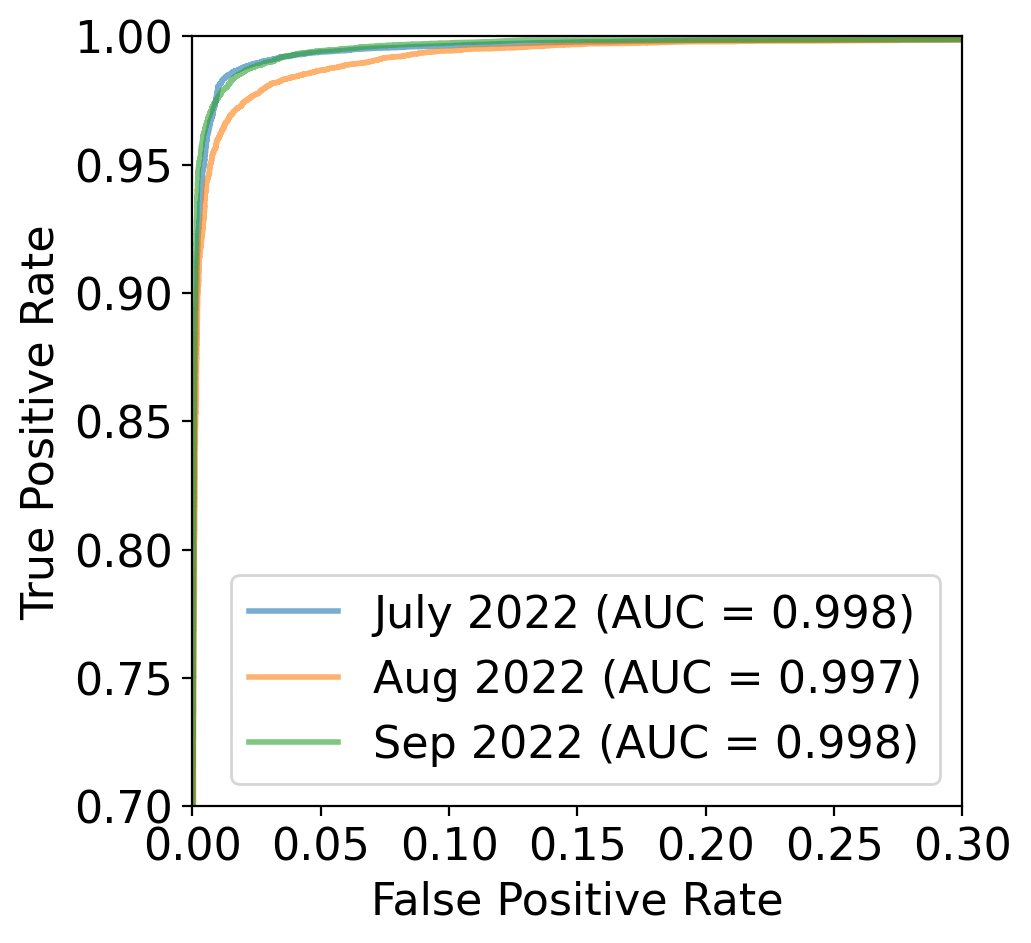}
    \caption{}
    \label{6roc:testing}
  \end{subfigure}\hfill
  \caption{(a) Validation, (b) Testing ROCs for meta-learner.}
  \label{6fig:rocs}
\end{figure}

\textbf{Model Training:} We empirically identify that semi-supervised graph learning followed by supervised classification yields favorable results compared to the models used for blocklist generation. Further, instead of utilizing the training data for a longer period (e.g. one month) to train a single model, having multiple models for different time slices and ensembling them yields superior results. As evaluated in Section~\ref{ss:metaanalysis}, we empirically identify that assembling 4 GNN encoders yields the best classification performance. We also identify the Random Forest as the best meta-learner classifier. 
For training, we perform passive DNS expansion on the collected ground truth and combine them with the 4 training graphs corresponding to the 4 GNN models. Then, we obtain 4 embeddings for each domain and the concatenated encoding is fed to the downstream Random Forest meta-learner to classify each domain as benign or malicious.

\textbf{Model Validation and Testing.} Fig.~\ref{6fig:rocs} and Table~\ref{tab:metavaltest} show the ROC curves and performance results for validation and testing datasets. Our model achieves consistently high AUC of 0.998 for different months for both validation and testing. 

\begin{table}[!ht]
\centering
\caption{Meta-learner performances \review{at 0.5\% FPR thresh.}\eat{ \nabeel{Better if we can show performance at 0.1\% as mentioned above}}}
\footnotesize
\begin{tabular}{l|c|c|c|c}
\toprule
&\textbf{Dataset} & \textbf{F1} & \textbf{Precision} & \textbf{Recall} \\ 
\midrule
\multirow{3}*{Validation}
    &Jul 2022 & 0.963 & 0.991 & 0.937 \\ 
 &Aug 2022 & 0.959 &0.991 & 0.929 \\ 
 &Sep 2022 & 0.933 & 0.984 & 0.887 \\ 
  \midrule
\multirow{1}*{Testing}
  & Oct 2022 & 0.968 & 0.995 & 0.943 \\ 
 \bottomrule
\end{tabular}
\label{tab:metavaltest}
\end{table}

\section{Design Choices and Classifier Analysis}~\label{sec:classifier}
\vspace{-3mm}
\fatihh{\subsection{Selection of Features and ML Approaches}}
\eat{If several models have similar classification performance, it is recommended to select the least complex model as it is easier to not only use in practice but also to debug if something goes wrong. To this end, we compare our proposed model with the state-of-the-art ML approaches.}
In our quest to determine the best-performing model and feature sets, Table~\ref{tab:results_baseline} presents the results for three different feature sets using the five different ML approaches used by state-of-the-art methods:

\begin{itemize}[leftmargin=*]
\itemsep0em
    \item Feature-based supervised learning: The classical machine learning approach, as employed in various studies~\cite{Notos_Antonakakis2010, bilge:2014:Exposure,  lexical2015, page:2019:mal, practicalattacks:SP:2024}, involves tabulating features and applying classical supervised learners. As representative algorithms, we employ Random Forest~\cite{Leistner:2009:SSLRandomForest}, XGBoost~\cite{chen2016xgboost}, LightGBM~\cite{ke2017lightgbm}, known for their superior performance.
    \item Label propagation: Has been applied in various studies~\cite{polonium:SIAM:2011, nazca:NDSS:2014, BPPhishingCCS:2022, bp_mal2:2020, marmite:CCS:2017}, iteratively assigns labels to unlabeled nodes based on seed node labels in their neighborhoods. In our research, we adopt one of the most promising label propagation approaches, namely belief propagation~\cite{Yedidia:2003:BP}. 
    \item Shallow embedding (unsupervised) + supervised learning: Shallow embedding approaches learn a unique embedding for each node based on the graph topology. We use Node2Vec \cite{node2vec:sigkdd:2016}, a popular transductive shallow embedding generator. We concatenate the generated embeddings with features and perform the downstream classification using a feature-based supervised model. 
    \item Deep embedding (unsupervised) + supervised learning: We employ an inductive unsupervised GraphSAGE~\cite{graphsage:nips:2017} to create deep embeddings for the domain nodes, followed by feature-based supervised classification.
    \item Semi-supervised GNN: We employ a semi-supervised GraphSAGE to classify unlabeled nodes based on the graph structure and node features.
\end{itemize}

With no domain features, deep embedding outperforms the other methods. However, all models result in a high FPR ranging from 12.8\% to 21.1\%. This shows that the network structure alone is insufficient to detect malicious domains. This is consistent with many malicious domains being hosted on shared hosting infrastructures where benign domains are also hosted. Thus, one needs distinct features to effectively differentiate between malicious and benign domains within such infrastructures. In the second set of experiments, we observe that all six models, which utilize lexical features only, exhibit improved performance compared to the experiments with no features. We attribute this improvement to the tendency of attacks to create lexically similar domains in the same infrastructure. In the third set of experiments focusing on hosting features, we observe that the models demonstrate superior classification performance compared to those without features and those utilizing only lexical features. Since, unlike benign domains, many malicious domains do not have consistent traffic and the hosting features capture these differences to differentiate between the two classes. The last set of experiments utilizes both lexical and hosting features. Not surprisingly, we obtained the overall best result compared to all the categories. Out of all the models, both GNN models (deep embedding and semi-supervised GNN) achieve high classification performance as they take into account both topology and node features simultaneously to learn to discriminate embeddings for the two classes. Out of the two GNN models, we consistently get slightly better classification performance for the semi-supervised model and thus utilize it for daily blocklist generation. \ccsupdate{You can find further details on our GNN hyperparameter search in Appendix~\ref{app:differentgnn}. Based on the grid search results we use GNN that comprises three layers with embedding dimensions of 256, employ a final layer that aggregates all embeddings from the preceding layers. }

\begin{table}
\centering
\caption{\centering Performance comparison of various ML approaches and feature sets \textcolor{black}{at 0.5 classification threshold}.}
\footnotesize
\resizebox{\columnwidth}{!}{
\begin{tabular}{c|c|c|c|c|c}
    \toprule
     & \multirow{2}{*}{\textbf{Model Type}} & \multicolumn{4}{c}{\textbf{Metrics}} \\ 
    \cline{3-6} 
    & & \textbf{Accuracy} & \textbf{Precision} & \textbf{Recall} & \textbf{FPR} \\
    \midrule
    \multirow{4}{*}{\rotatebox[origin=c]{90}{Without}}
    \multirow{4}{*}{\rotatebox[origin=c]{90}{Features}}
    & Belief Propagation & 0.757 & 0.711 & 0.715 & 0.211  \\
    & Node2Vec & 0.807 & 0.801 & 0.798 & 0.148   \\
    & Unsup. GNN & 0.871 & 0.842 & 0.872 & 0.128  \\
    & Semisup. GNN & 0.815 & 0.767 & 0.815 & 0.185  \\
    \midrule
    \multirow{7}{*}{\rotatebox[origin=c]{90}{Lexical}} 
    \multirow{7}{*}{\rotatebox[origin=c]{90}{Features}} 
    & Random Forest & 0.864 & 0.880 & 0.801 & 0.085  \\
    & LightGBM & 0.868 & 0.897 & 0.790 & 0.071  \\
    & XGBoost & 0.861 & 0.882 & 0.788 & 0.082  \\
    & Node2Vec & 0.909 & 0.933 & 0.855 & 0.048  \\
    & Unsup. GNN & 0.913 & 0.927 & 0.871 & 0.053  \\
    & Semisup. GNN & 0.938 & 0.946 & 0.904 & 0.037  \\
    \midrule
    \multirow{7}{*}{\rotatebox[origin=c]{90}{Hosting}}
    \multirow{7}{*}{\rotatebox[origin=c]{90}{Features}}
    & Random Forest & 0.944 & 0.958 & 0.914 & 0.031  \\
    & LightGBM & 0.945 & 0.959 & 0.915 & 0.030  \\
    & XGBoost & 0.944 & 0.957 & 0.914 & 0.032  \\
    & Node2Vec & 0.953 & 0.962 & 0.930 & 0.028 \\
    & Unsup. GNN & 0.968 & 0.971 & \textbf{0.954} & 0.021  \\
    & Semisup. GNN & 0.967 & 0.970 & 0.951 & 0.021  \\
    \midrule
    \multirow{7}{*}{\rotatebox[origin=l]{90}{Lex. \& Hosting}}
    \multirow{7}{*}{\rotatebox[origin=c]{90}{Features}}
    & Random Forest & 0.952 & 0.971 & 0.920 & 0.021  \\
    & LightGBM & 0.956 & 0.974 & 0.925 & 0.019  \\
    & XGBoost & 0.957 & 0.971 & 0.931 & 0.021  \\
    & Node2Vec & 0.962 & 0.977 & 0.935 &  0.017 \\
    & Unsup. GNN & 0.965 & 0.970 & 0.949 & 0.022  \\
    & Semisup. GNN & \textbf{0.982} & \textbf{0.984} & 0.941 & \textbf{0.012}  \\
    \bottomrule
\end{tabular}
}
\label{tab:results_baseline}
\end{table}

\subsection{PDNS Expansion}
\label{subsec:pdnsexpansion}

\begin{figure}
\centering
\includegraphics[width=0.80\columnwidth]{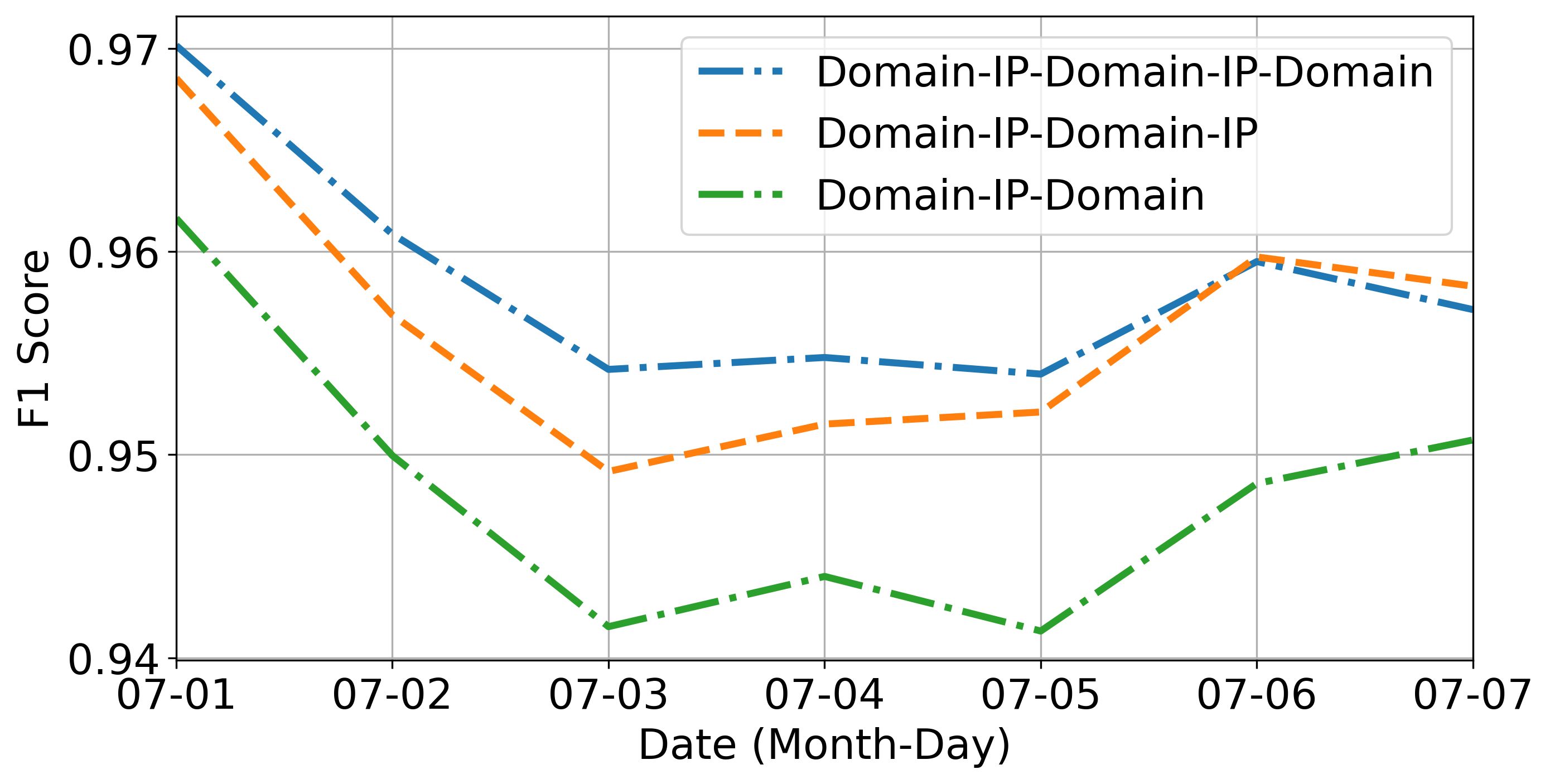}
\caption{Graph structure comparison. \eat{\ik{Make "Mantis" capital in the legend}} \eat{\nabeel{Fatih: Would you be able to repeat the experiment for another level? i.e.. D-IP-D-IP-D. One possible issue is the scale of the graph.}}}
\label{fig:expansion_types}
\end{figure}

\begin{figure}
\centering
\begin{subfigure}[t]{0.45\linewidth}
    \includegraphics[width=\linewidth]{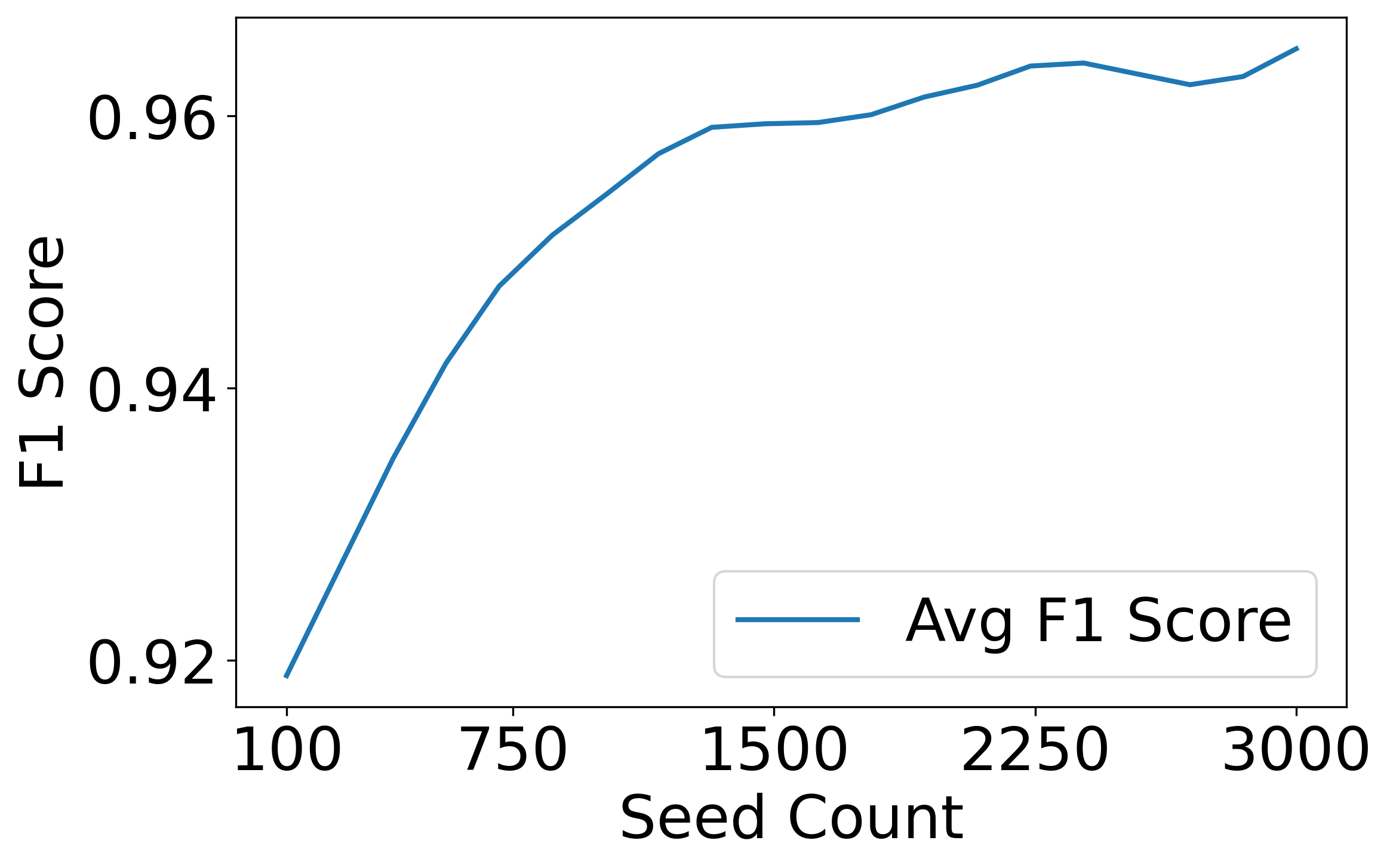}
    \caption{}
    \label{fig:exp_trainingsizefpr}
  \end{subfigure}
\begin{subfigure}[t]{0.45\linewidth}
    \includegraphics[width=\linewidth]{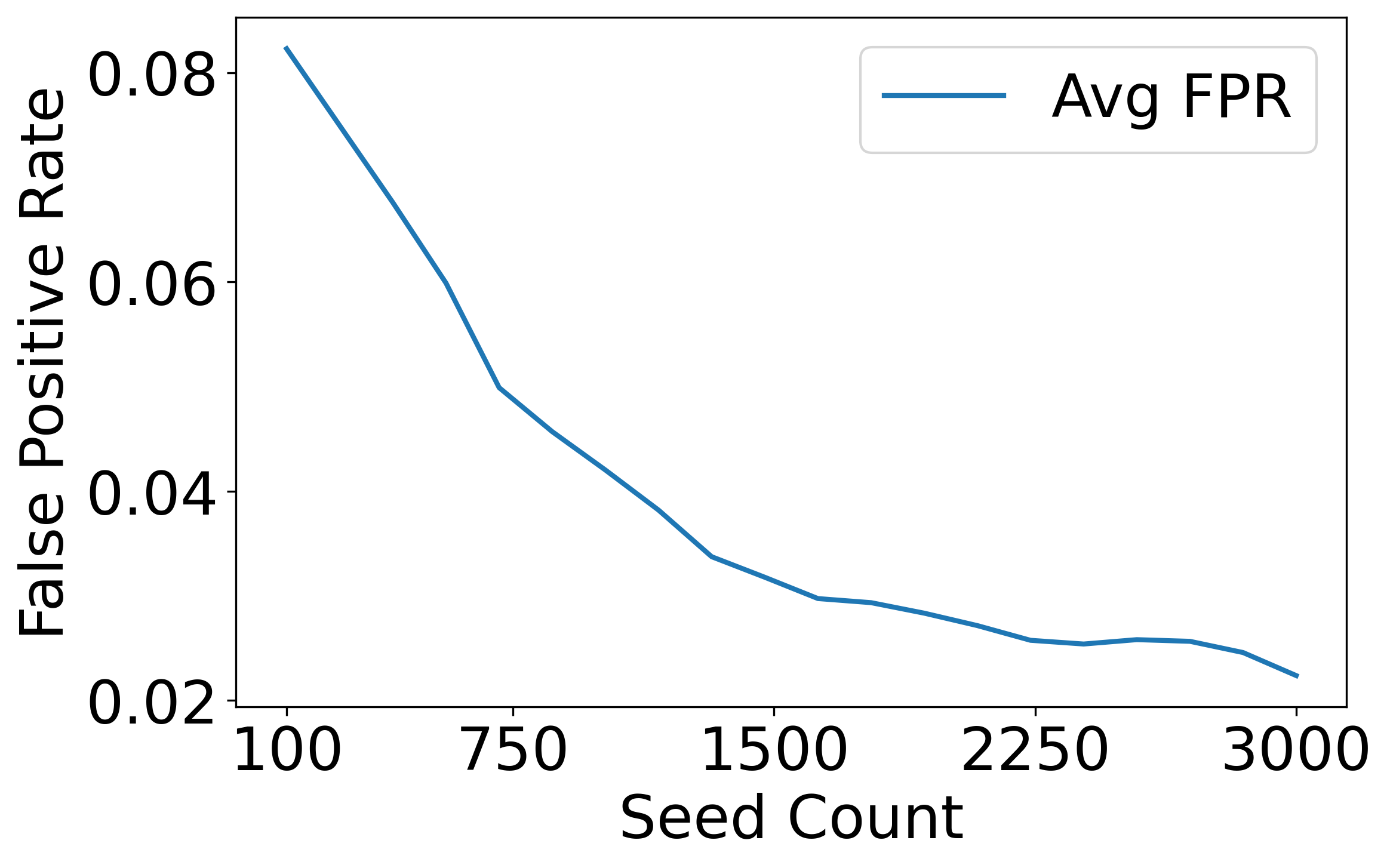}
    \caption{}
    \label{fig:exp_trainingsizeacc}
  \end{subfigure}  \hfill
  \begin{subfigure}[t]{0.45\linewidth}
    \includegraphics[width=\linewidth]{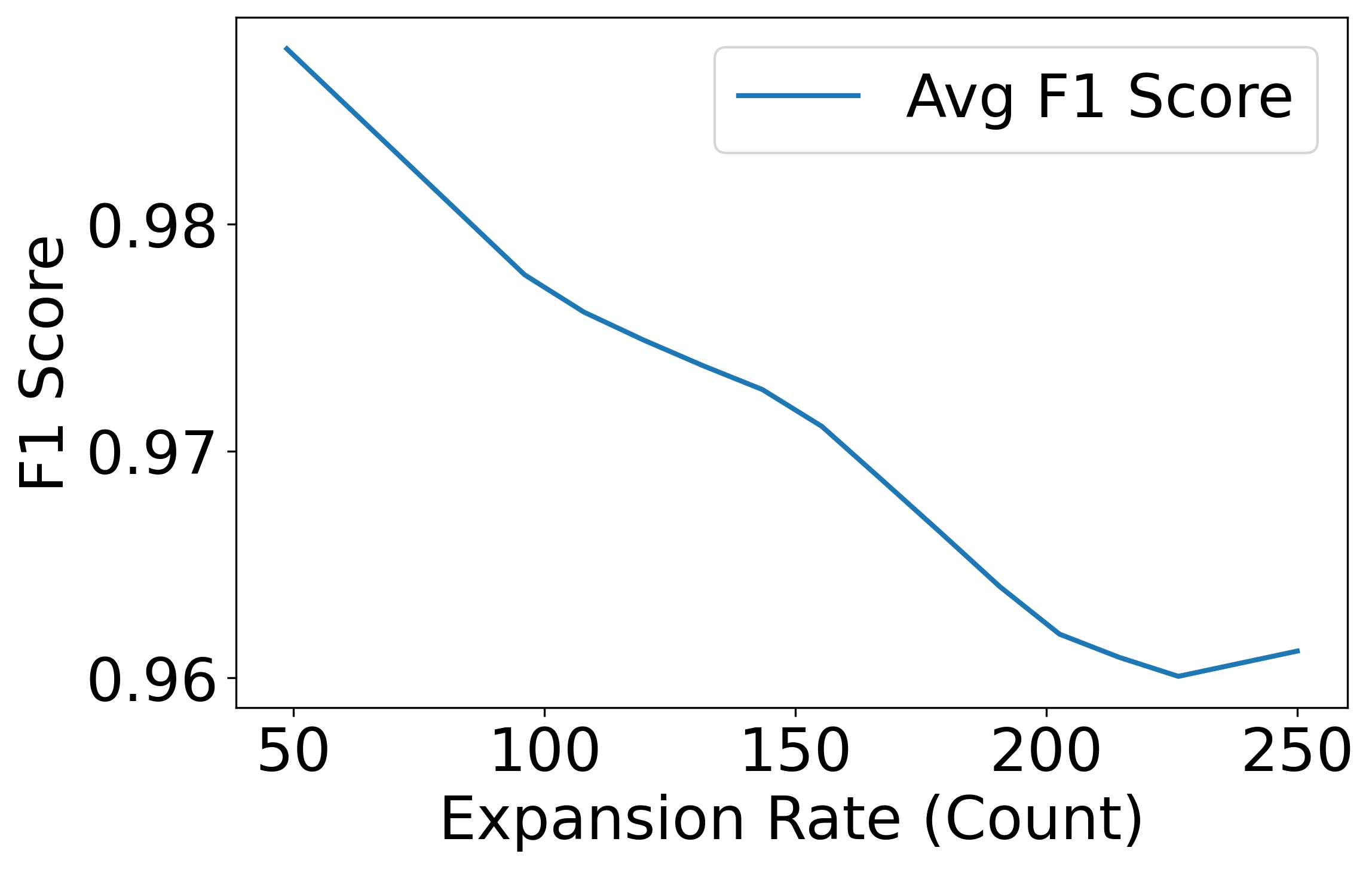}
    \caption{}
    \label{fig:exp_expansionrateacc}
  \end{subfigure}
  \begin{subfigure}[t]{0.45\linewidth}
    \includegraphics[width=\linewidth]{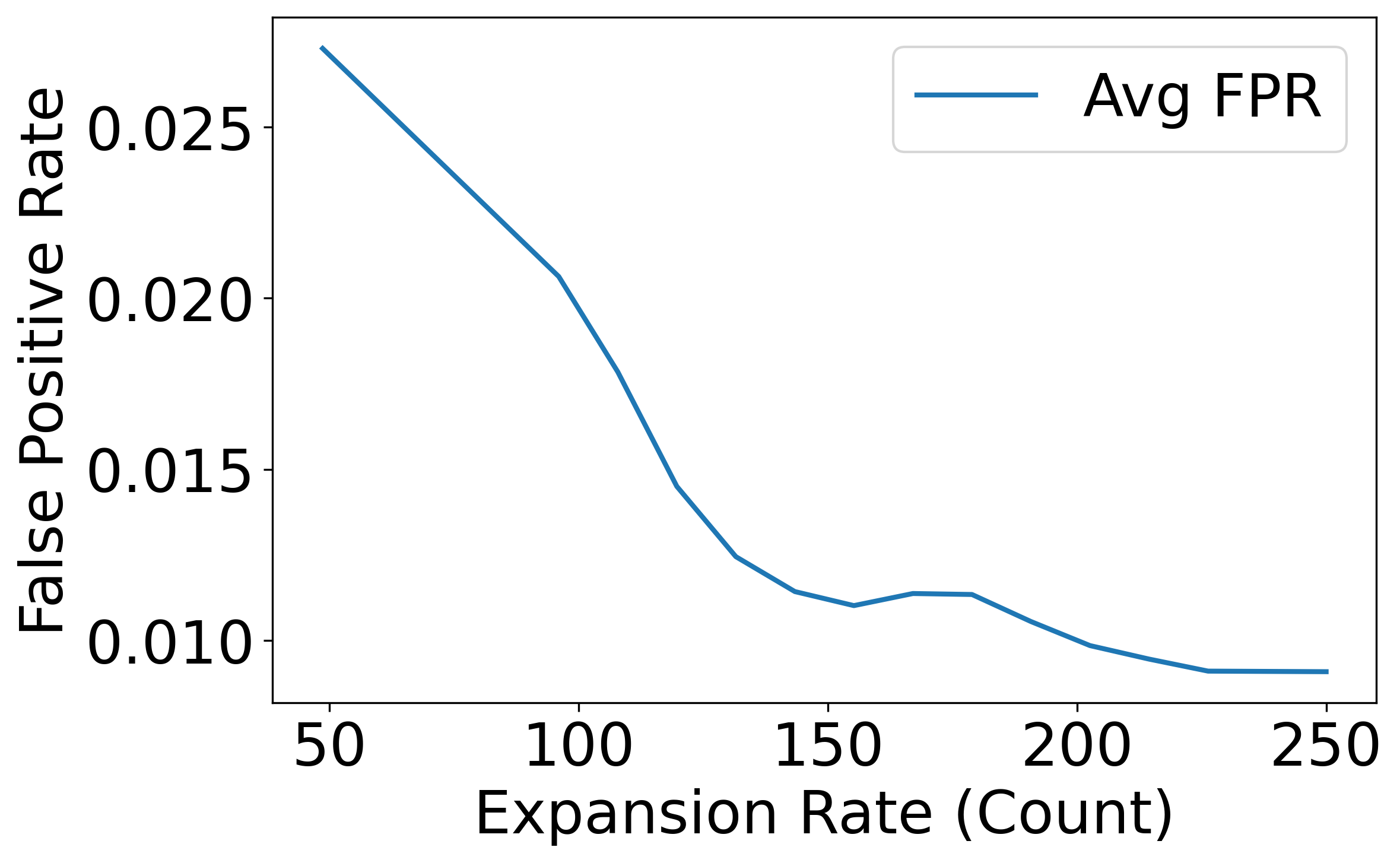}
    \caption{}
    \label{fig:exp_expansionratefpr}
  \end{subfigure}\hfill
    \caption{(a) F1 score, (b) FPR for different seed domain counts. (c) F1 score, (d) FPR for different expansion rates.}
  \vspace{-2mm}
  
  \label{fig:exp_trainingsize}
\end{figure}

In this section, we empirically estimate the impact of two key parameters of the PDNS expansion algorithm: the number of hops and the number of recently hosted domains considered in each hop. 
We perform Level 1 (Domain-IP-Domain), Level 2 (Domain-IP-Domain-IP), and Level 3 (Domain-IP-Domain-IP-Domain) expansions for each day from July 1st to 7th, 2022. Fig.~\ref{fig:expansion_types} shows the F1-score for each day for different expansion levels. We observe that the Level 2 expansion consistently yields a 1\% improvement compared to Level 1. However, the gain in F1-score for Level 3 is less than 0.4\% in general. As Level 3 domains are farther away from the seed nodes, they have less influence from the seed domains. Additionally, the graph size exponentially increases from Level 2 (around 0.5 million nodes on average) to Level 3 (around 3 million nodes on average), incurring a huge computational cost. Based on these findings, we empirically fix the graph expansion to Level 2.
While keeping the expansion to Level 2, we perform experiments to identify the optimal number of recent domains hosted on each IP (i.e. expansion rate) that yields a high F1-score and a low FPR. Fig.~\ref{fig:exp_expansionrateacc} and~\ref{fig:exp_expansionratefpr} show the F1-score and FPR for different expansion rates from 50 to 250, respectively. It shows that as the expansion increases, F1-score falls slowly while FPR falls rapidly and plateaus at 200-250. Since our primary goal is to minimize false positives and reduce the burden on security operations teams, we set the expansion rate to 200 in our experiments. 

\subsection{Daily Blocklist and Training Window Size}
\label{subsec:datasize}

Collecting labeled training data is an expensive and timely process. Therefore, it is important to identify the minimum training data size for the best empirical performance. \spfinal{Fig.~\ref{fig:exp_trainingsize} shows the performance for different seed domain counts from 100 to 3000 (i.e., different training data sizes), averaged over three temporally disjoint graphs. We observe that the F1-score rapidly increases as the seed size increases and starts to plateau when the seed size reaches 2000. Based on this experiment, we recommend using at least 2,000 seed domains for optimal results with high recall and very low false positive rate. In our daily blocklist generation pipeline, we employ on average 3,000 domains per day. However, it should be noted that one may use our approach with small seeds with only hundreds of malicious domains and still achieve an F1-score above 90\% at the expense of slightly lower recall. One needs to select different classification thresholds depending on the seed size and the desired false positive rate.}

As explained in Section~\ref{subsec:pdnsexpansion}, we generate a daily passive DNS graph, and the window size determines the number of consecutive days combined in each graph. This feature sheds light on the duration of hosting infrastructure reuse by attackers, as empirically analyzed in Fig.~\ref{fig:overlapping_ip_july}. By combining daily graphs, we add more connections to the leaf IP nodes, consequently enhancing the overall performance of our trained models. Our model is trained using data from different window sizes ranging from 1 to 15 days. 
Notably, as the window size increases, FPR gradually converges to a value closer to 1\%, particularly from day 7 onwards. 
Considering the graph size (i.e. computational cost detailed in Appendix~\ref{app:differentgnn}) and FPR, we empirically fix our window size to 7 days.

\eat{To ensure consistently high performance, we train \system with different training data sizes and try to find the best performing one. Besides the size of the training window, which specifies the number of daily graphs to be combined, the sizes of individual graphs are affected by two factors: the number of seed domains and the rate of expansion around those domains. In this section, we perform the experiments on seed domain counts and evaluate PDNS expansion parameters in the following section. Fig.~\ref{fig:exp_trainingsize} shows the performance for different seed domain counts from 500 to 3000 for three temporally different graphs. We observe that seed count and \system's performance are positively correlated and in order to ensure consistently high performance one needs at least 2000 seed nodes. Considering our daily average seed count of 2877 domains and 7-day training window size, we use much more than 2000 seed nodes, on average, to train our models.}

\subsection{Generalizability of the Model and Predictive Performance over Time}
\label{subsec:generalizability}

\begin{figure}
\centering
\includegraphics[width=0.80\columnwidth]{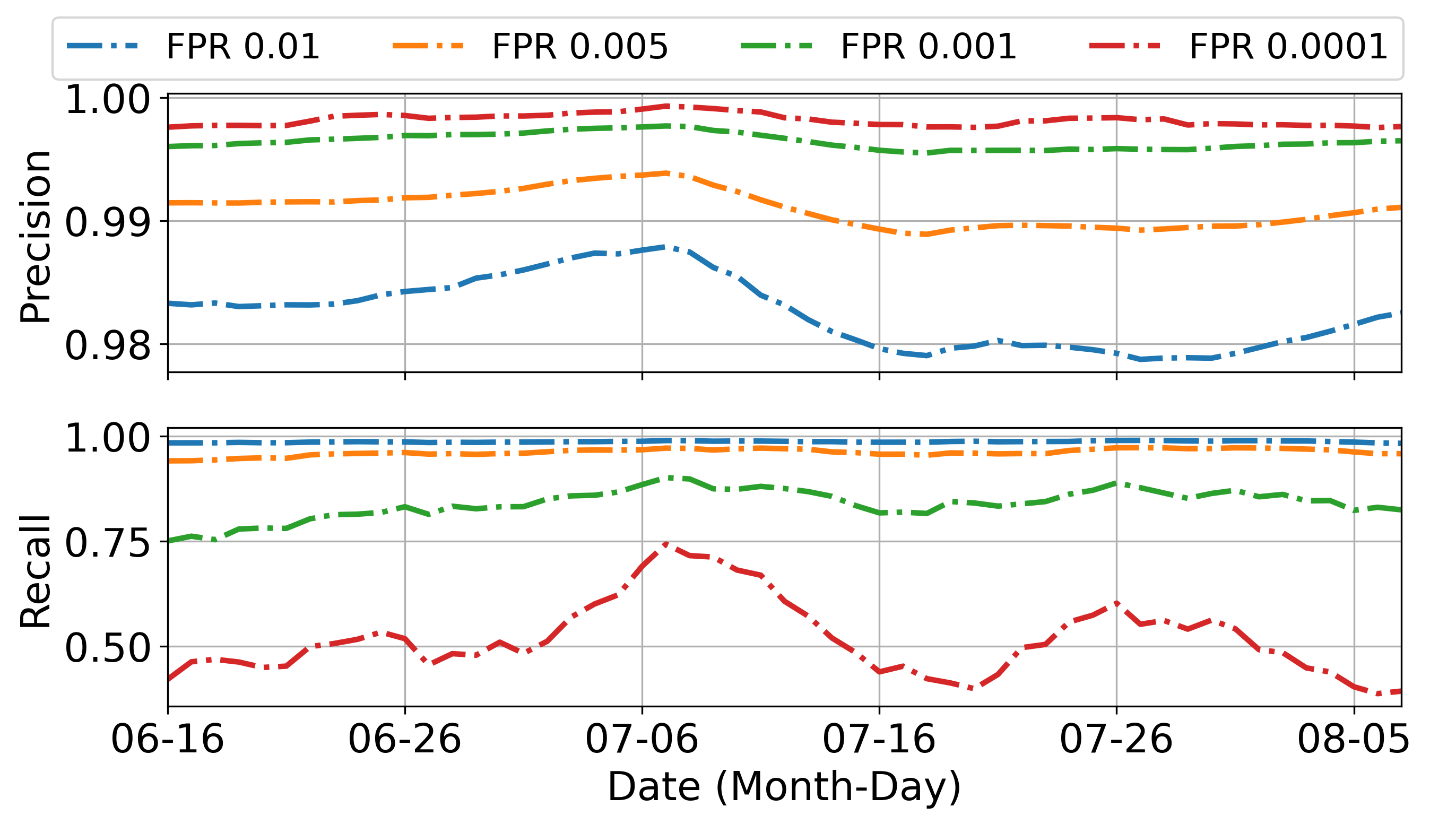}
\caption{Generalizability of the model.\eat{ across temporally different datasets from 2022-06-10 to 2022-08-10.}}
\label{fig:exp_logitivity}
\end{figure}

\review{Maintaining consistent high performance over temporally different datasets is crucial for practical machine learning models. Since its deployment, \system has consistently delivered outstanding performance. Fig.~\ref{fig:exp_logitivity} illustrates daily precision and recall metrics for different FPR values, each calculated within a 7-day training window. 
\system provides an excellent trade-off between FPR and recall and can substantially reduce FPR while making some trade-offs in recall: On average, \system can achieve FPR 0.5\%, with recall 95.6\%, FPR 0.1\% with recall 86.9\%.
}

\begin{figure}
\centering
\includegraphics[width=0.80\columnwidth]{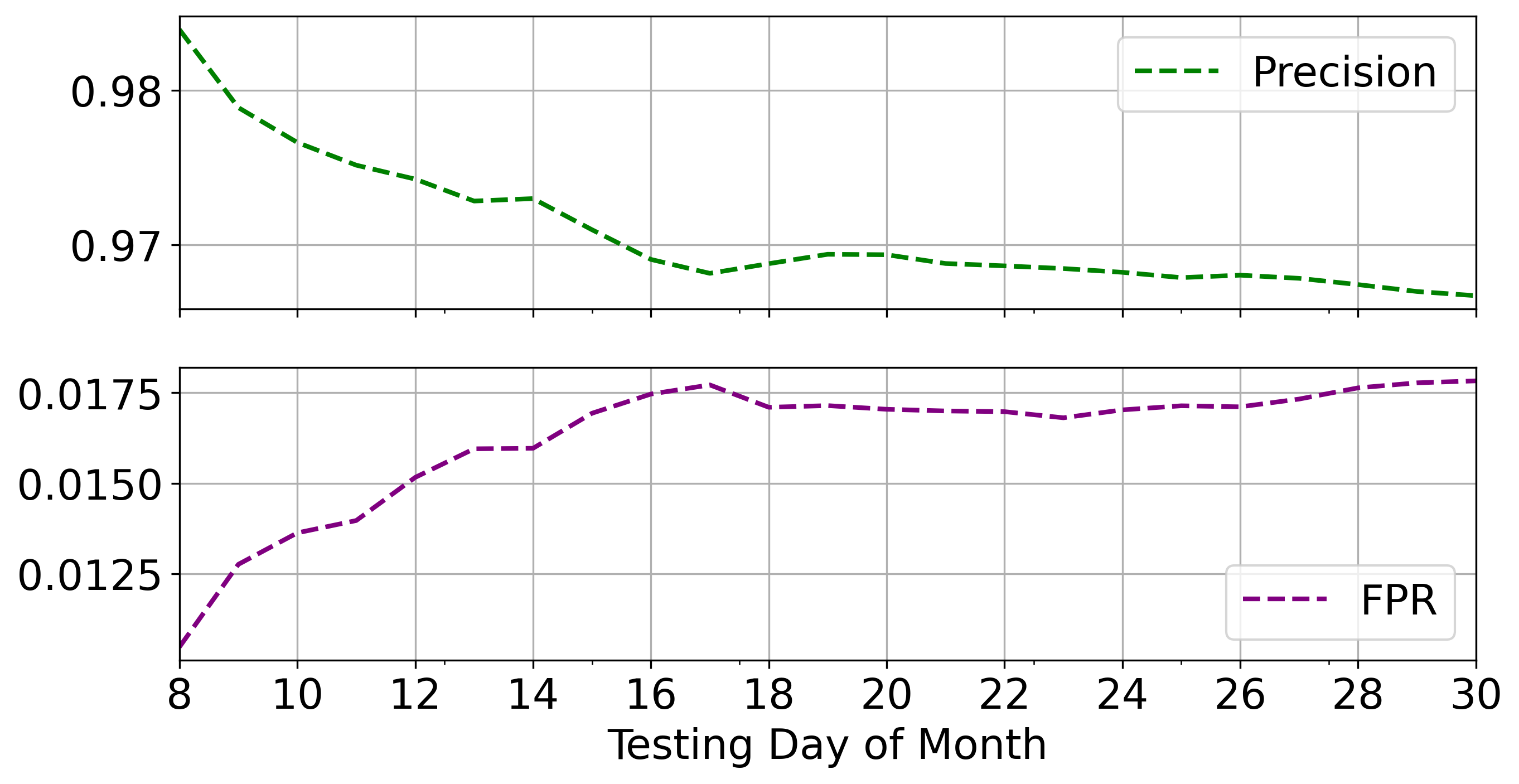}
\caption{Predictive performance of the model.}
\label{fig:predictiveperf}
\end{figure}

The ability to employ a trained model for a specific duration without the need for re-training offers several practical advantages, including reduced training time, optimized resource utilization, and lower labeling costs. To evaluate the predictive capability of our system, we train a model using one-week window data and predict unseen malicious domains in the days following the training window. This experiment is repeated at the beginning of three months, and as illustrated in Fig.~\ref{fig:predictiveperf} our system maintains its performance within a 3\% precision margin on average even after three weeks for unseen data. Given that the retraining cost falls within our budget and we prioritize achieving the highest precision feasible, our current deployment of the system opts for daily retraining. However, if daily retraining is not feasible, we recommend retraining the system at least bi-weekly, as the decline in precision becomes noticeable after 14 days.

\subsection{SoTA Comparison and Optimizations}\label{subsec:ablation}

\ccsupdate{
We compared our approach with the recent state-of-the-art methods. 
\spupdate{Galloway et al.~\cite{practicalattacks:SP:2024} integrate different feature sets with features extracted from the network structure to train a Random Forest model.}
Nabeel et al.~\cite{bp_mal2:2020} apply belief propagation and Kim et al.
~\cite{BPPhishingCCS:2022} employ loopy belief propagation. Other graph-based approach ~\cite{ringer:ICCS:2020, handom:wang:2023, blocklist_raid2024} leverage the power of GNNs on graphs containing client information. Ringer~\cite{ringer:ICCS:2020} employs a dynamic GCN model with lexical features and calculates attention weights based on the shared neighborhood, while HanDom~\cite{handom:wang:2023} utilizes an attention-based HAN model with time-based features. Blocklist-forecast~\cite{blocklist_raid2024} builds a heterogeneous graph and uses HinSAGE embeddings along with a Random Forest classifier. We implemented these approaches to the best of our knowledge, and their optimal results on our datasets are showcased in Table~\ref{tab:sota}. The intuition behind existing graph-based approaches is that a client associated with a malicious domain is likely to have connections with other malicious domains. However, our results suggest that these approaches are optimized for benign environments. As the results indicate, by focusing on recent attacks, we can efficiently identify other attacks.

\begin{table}
\centering
\caption{\centering Performance comparison with SoTA Methods.}
\footnotesize
\resizebox{\linewidth}{!}{
\begin{tabular}{l|l|c|c|c|c}
    \toprule
        &\textbf{Model} & \textbf{F1} & \textbf{Prec.} & \textbf{Recall} & \textbf{FPR} \\ 
    \midrule
        Nabeel et. al.~\cite{bp_mal2:2020} & BP & 71.3 & 71.1 & 71.5 &  21.1 \\
    Kim et. al.~\cite{BPPhishingCCS:2022} & BP & 79.1 & 80.4 & 77.8 &  17.8 \\
    Galloway et. al.~\cite{practicalattacks:SP:2024} & RF & 92.4 & 92.5 & 92.3 &  8.7 \\     Liu et. al.~\cite{ringer:ICCS:2020} & GNN & 94.0 & 95.5 & 92.6 &  3.8 \\     Wang et. al.~\cite{handom:wang:2023} & GNN & 94.8 & 96.9 & 92.8 &  3.4 \\ 
        Kumarasinghe et. al.~\cite{blocklist_raid2024} & GNN & 94.3 & 94.2 & 94.4 & 5.9 \\
    \system (ours) & GNN & \textbf{97.3} & \textbf{99.0} & \textbf{95.6} & \textbf{0.5}  \\
    \bottomrule
\end{tabular}
}
\label{tab:sota}
\end{table}

We further conducted an experiment to validate our decision to distinguish between attacker-owned and compromised domains, and exclude web hosting domains.
By making such decisions, we achieve higher precision (99.0\%) and recall (95.6\%) with a lower FPR of 0.5\%. In contrast, when compromised and webhosting domains are included: lower precision (93.69\%) and recall (89.80\%) with a higher FPR (4.55\%), reflecting a notable decrease of 4\% in each metric. 
Note that even in this setting, our approach still outperforms recent state-of-the-art approaches such as~\cite{BPPhishingCCS:2022} and~\cite{practicalattacks:SP:2024}.
In addition to assessing the implications of our decisions, we conducted a thorough analysis of our computational performance, which can be found in Appendix~\ref{app:differentgnn}.
}

\subsection{Robustness}\label{subsec:robustness}

\begin{figure}[t]
\centering
\includegraphics[width=0.8\columnwidth]{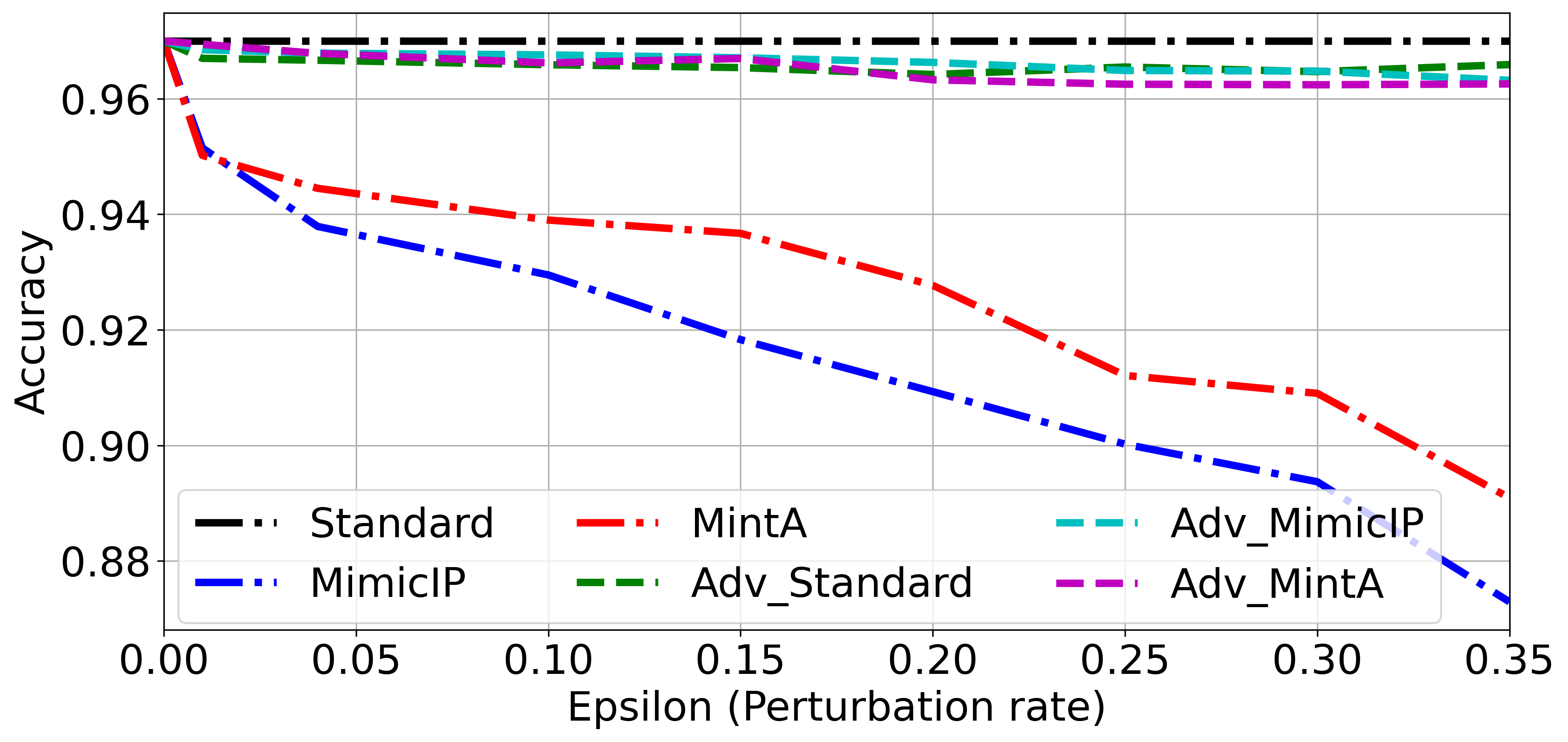}
\caption{Robustness of standard and adversarially trained models.}
\label{fig:adversarial}
\end{figure}

\spfinal{
Adversarial ML techniques craft samples of a particular class to evade detection. 
Most research directly manipulates the feature space or edge weights to craft samples of a particular class. However, these techniques are not practical especially in the cybersecurity domain when they cannot be converted to realistic input samples. In this section, we evaluate our approach against recent SoTA practical DNS attacks~\cite{practicalattacks:SP:2024, minta:SP:2024}. 
Galloway et al.~\cite{practicalattacks:SP:2024} introduces attacks targeting graph structure, and popularity- and registration-based features. While registration-based attacks are shown to have minimal impact on a domain's reputation, attacks on graph structure and popularity-based features were found to be effective. We believe using popularity-based features and ground truth together potentially exaggerate the impact of Galloway et al.'s most successful attack. Since we intentionally omit popularity rankings from our feature sets, our approach remains resilient against most successful popularity list manipulation attacks and thus we only evaluate our approach against MimicIP, the most effective graph-based attack proposed in Galloway et. al.~\cite{practicalattacks:SP:2024}.\eat{This is the only practical attack among the proposed attacks that does not have a financial cost for the attacker.}} In this attack, the attacker inserts A records resolving to IPs associated with benign domains listed in the Tranco Top 100K list. As we consider all historical resolutions, even if the attacker temporarily halts malicious activity, this results in additional edges between malicious domains and IPs linked to the popular domains. During training and testing, we greedily select $n$ IPs ($n=1,2,3$) that induce the highest increase in prediction score. 
\spfinal{
While MimicIP focuses on single-domain attacks that attempt to mimic benign domains, MintA~\cite{minta:SP:2024} addresses a scenario where the attacker controls multiple domains and has access to a surrogate model, allowing them to compute the loss for given graphs and strategically manipulate IP resolutions of controlled malicious domains.
We provide a comprehensive evaluation of our models on both standard and adversarial samples for both attacks. Additionally, we train a model using a combination of standard and adversarial samples and analyze its performance. Figure~\ref{fig:adversarial} shows the performance of our approach under both clean and adversarial conditions across different perturbation rates, where the perturbation rate represents the percentage of domains subjected to the specified attack. As expected, the performance of the model trained on standard samples decreases under the adversarial setting with an increasing perturbation rate. However, due to the way we construct the graph, the rate of degradation of performance is low and our approach still achieves over 90\% of accuracy even with a high perturbation rate of 15\%. Further, training our model with adversarial samples nearly results in the same original performance under the adversarial setting, further demonstrating the robustness of our approach.} To ensure reproducibility, we have included our training code for both scenarios in the repository.

\subsection{Meta-Learner Analysis}~\label{ss:metaanalysis}
In this section, we empirically evaluate our design choices with the meta-learner for the on-demand classifier.

\begin{table}[!ht]
\centering
\caption{On-demand classifier testing results. \eat{\nabeel{Let's create the domain dataset from GSB for Oct and add another row to this table. Also, another one for Tranco.}}}
\footnotesize
\label{tab:results_baseline_meta}
\resizebox{\linewidth}{!}{
\begin{tabular}{l|c|c|c|c|c}
    \toprule
    \multirow{2}{*}{\textbf{Dataset}} & \multirow{2}{*}{\textbf{\#Domain}} & \multicolumn{2}{c|}{\textbf{Baseline RF}} & \multicolumn{2}{c}{\textbf{On-Demand Classifier}} \\ 
    \cline{3-6} 
    & & \textbf{Accuracy} & \textbf{F1} & \textbf{Accuracy} & \textbf{F1} \\
    \midrule
    Alexa & 19767 & 96.8 & - & 99.1 & - \\ 
    Tranco & 10000 & 98.6 & - & 99.4 & - \\ 
    CrUX & 10000 & 98.4 & - & 99.6  & -\\ 
    \midrule
    VT & 19408 & 95.2 & 97.5 & 96.8  & 98.4\\ 
    GSB & 1000 & 73.6 & 84.8 & 94.1  & 97.0 \\ 
    OpenPhish & 2115 & 89.6 & 94.5 & 96.3  & 98.1\\ 
    PhishTank & 2215 & 93.1 & 96.4 & 97.1  & 98.5\\ 
    
    \bottomrule
\end{tabular}
}
\label{tab:dataset_real}
\end{table}

{\bf On-Demand Performance on Different Datasets:} We also measure \system's on-demand performance on testing datasets from different sources. 
To mitigate temporal bias, we perform these tests on previously unseen data collected from the following month.
In addition to VT, we also rigorously test our approach on diverse datasets, including OpenPhish, PhishTank, GSB, and benign datasets. 
\eat{Malicious domains are selected from the first seen, i.e. newly observed domains, feeds on October 2022 from the respective sources. }In order to have labels with high confidence, we select only manually verified URLs by PhishTank and OpenPhish and consider the top 100k domains from the top lists. 
\eat{Since the voting process used by those intelligence feeds could be compromised, we do sanity-checking on the voting results by actively querying VT and filtering out URLs that receive less than three positives.} Additionally, we actively query all the domains in VT and exclude those from malicious sources (PhishTank, OpenPhish, VT) that are not marked by VT as malicious, and exclude those from the benign sources marked as malicious. Our evaluation results, as presented in Table~\ref{tab:dataset_real}, consistently demonstrate high classification performance and inductive nature of our approach. 

\begin{figure}
\centering
\begin{subfigure}[t]{0.49\linewidth}
    \includegraphics[width=\linewidth]{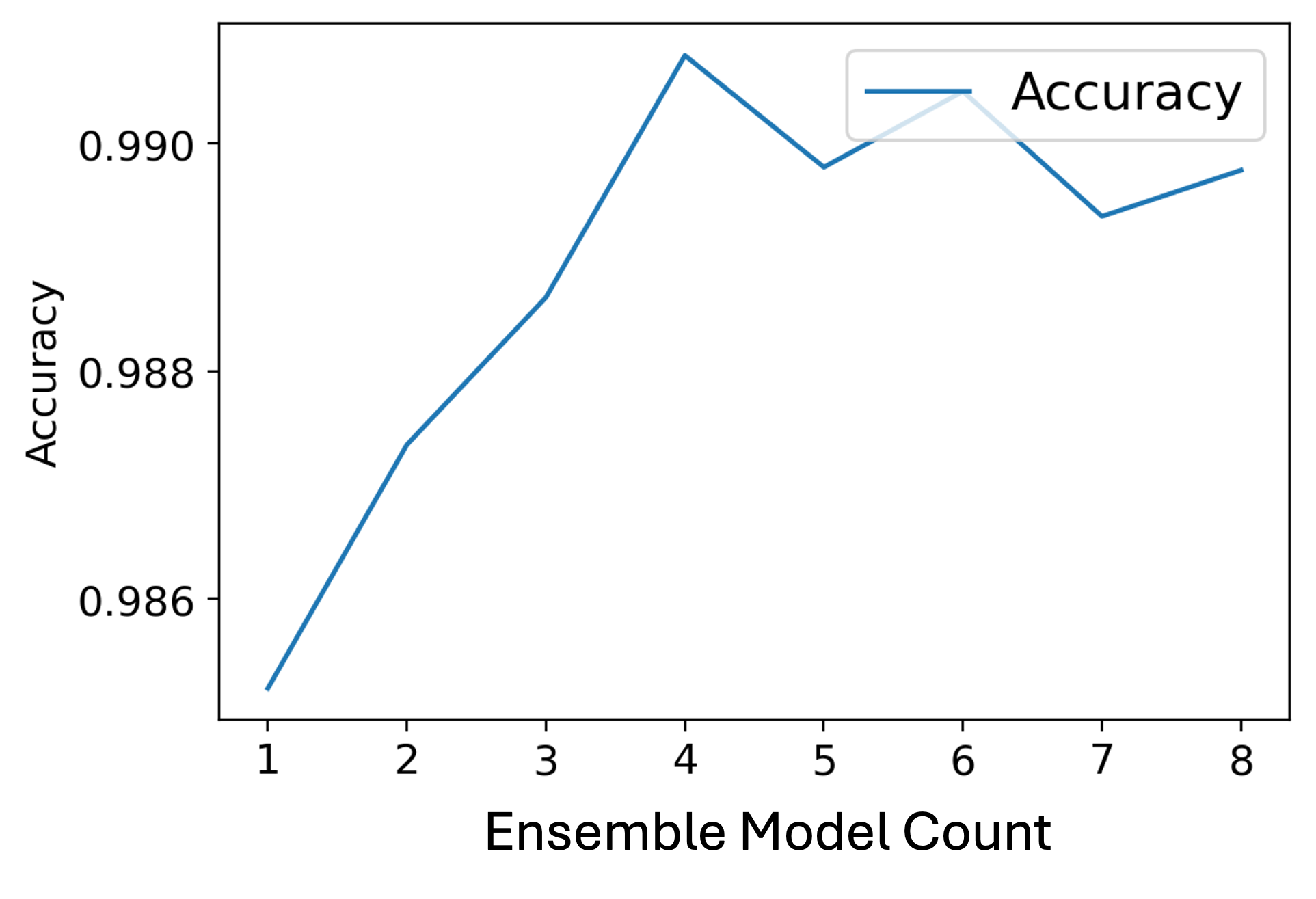}
    \caption{}
    \label{fig:exp_ensemblecountacc}
  \end{subfigure}
  \begin{subfigure}[t]{0.49\linewidth}
    \includegraphics[width=\linewidth]{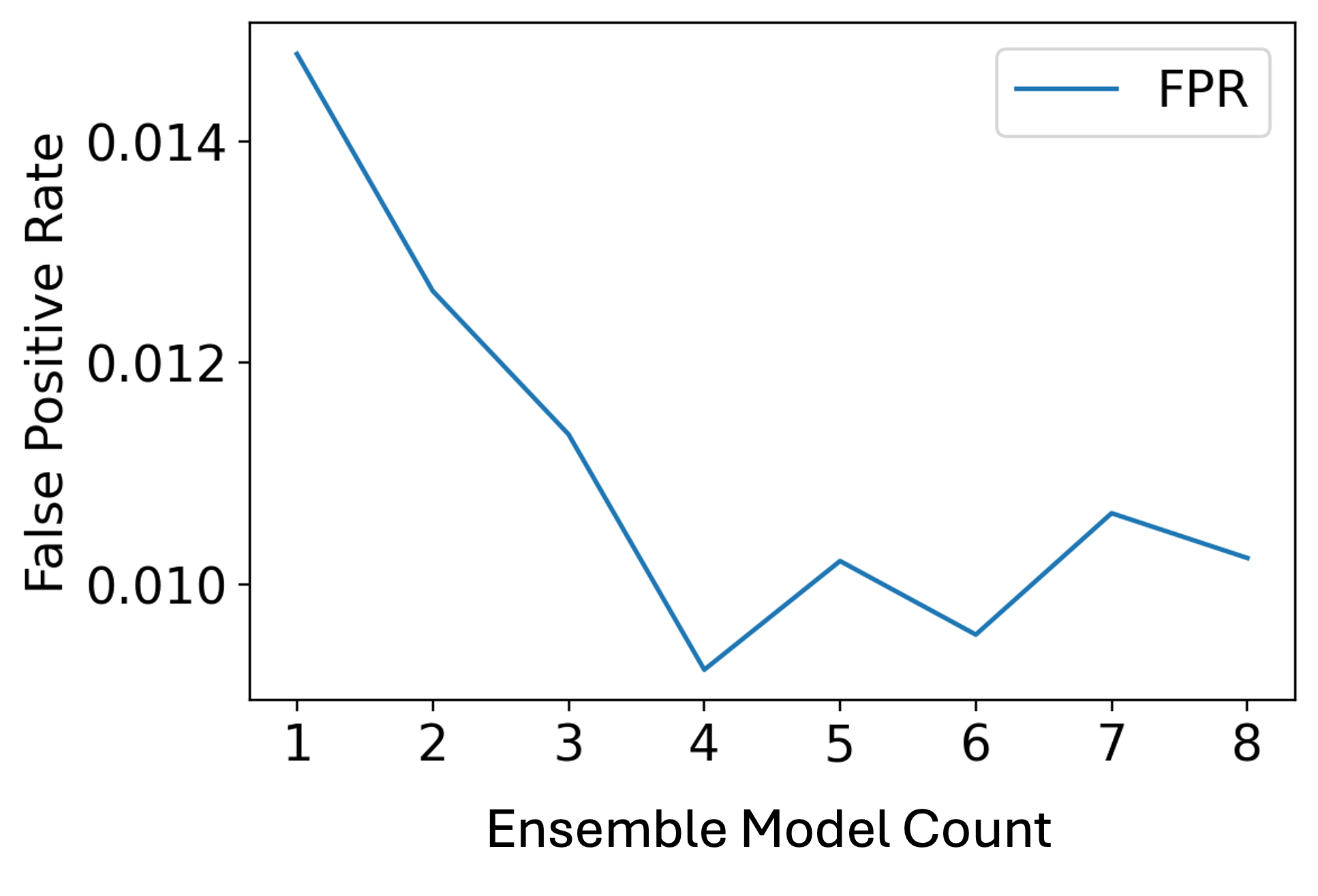}
    \caption{}
    \label{fig:exp_ensemblecountfpr}
  \end{subfigure}\hfill
  \caption{(a) Acc., (b) FPR for different ensemble counts.}
    \label{fig:exp_ensemblecount}
\end{figure}

{\bf Number of Ensemble Models:} An important aspect of the meta-learner is to identify the best number of ensemble models. We measure the accuracy and FPRs for varying numbers of ensemble models on the Alexa top 100K domains using non-overlapping weekly trained GNN models from  2022-07-01, and 2022-09-30. 
As shown in Fig.~\ref{fig:exp_ensemblecount}, the performance improves as the number of models increases and peaks at 4 models. Hence, we fix the number of GNN encoders in our on-demand classifier to 4. 

\section{Post Analysis}~\label{sec:post}
\vspace{-4mm}

\begin{figure}[t]
\centering
\includegraphics[width=0.85\columnwidth]{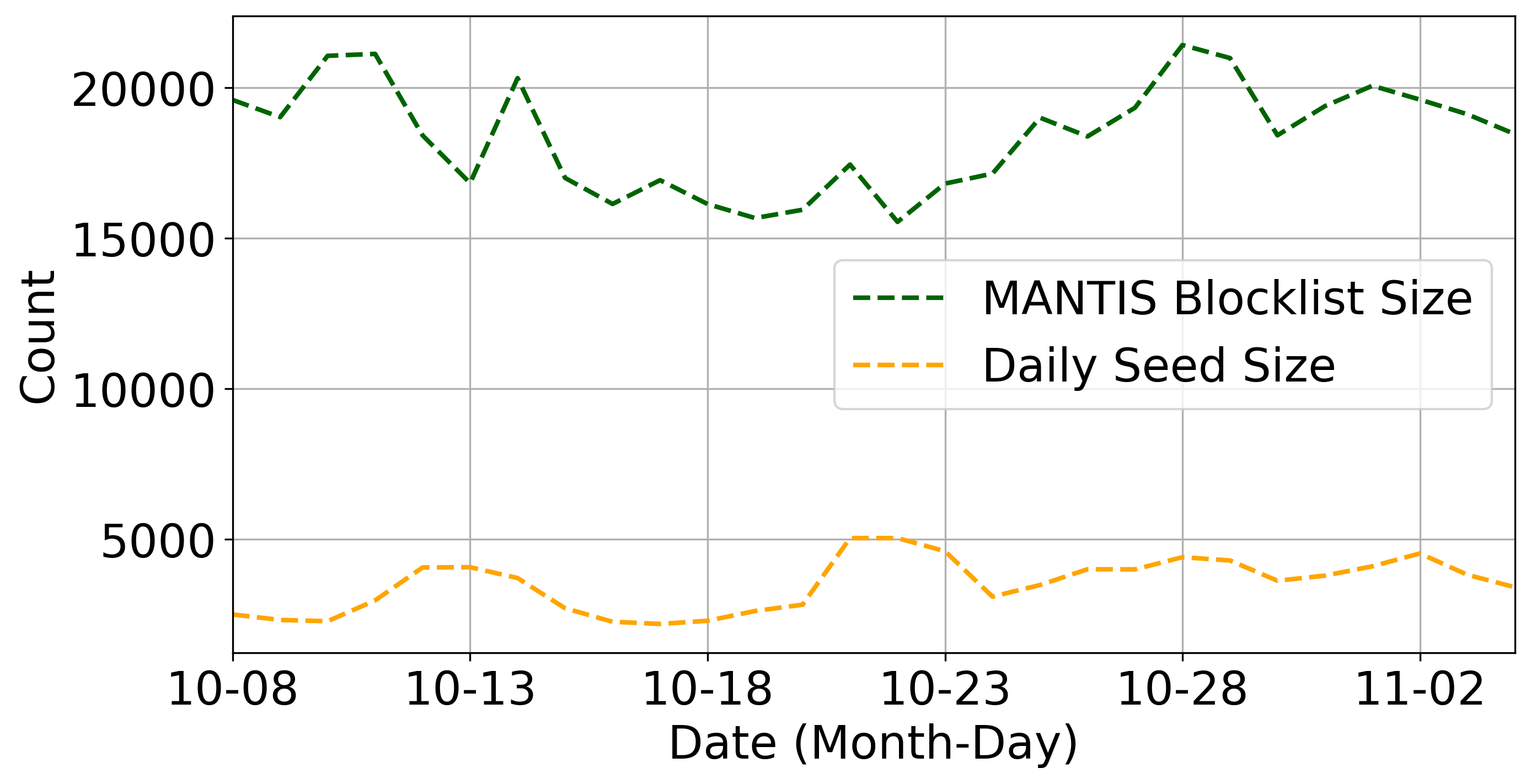}
\caption{Number of newly detected malicious domains.\eat{\fatih{For this figure count ratio MANTIS/VT=5.20.}}}
\vspace{-3mm}
\label{fig:dailyblocklistcount}
\end{figure}

For the daily blocklist generation, we set the FPR to 0.1\% and apply additional filtering to exclude potentially impactful benign domains, in line with common cybersecurity practices that combine rule-based methods with ML models~\cite{mink:SP:2023}. Our manual spot-checking of random samples (Section~\ref{subsec:analysisofdetectedmalicious}), confirms the validity of this measurement and underscores the representativeness of our ground truth.
Fig.~\ref{fig:dailyblocklistcount} shows the seed domain count and the number of newly predicted malicious domains at an FPR of 0.1\%. On average, \system detects over 5 new malicious domains for each seed malicious domain.

\subsection{GNN Explanations}
\label{sec5:feature_importance}
\begin{figure}
\centering
\includegraphics[width=0.99\columnwidth]{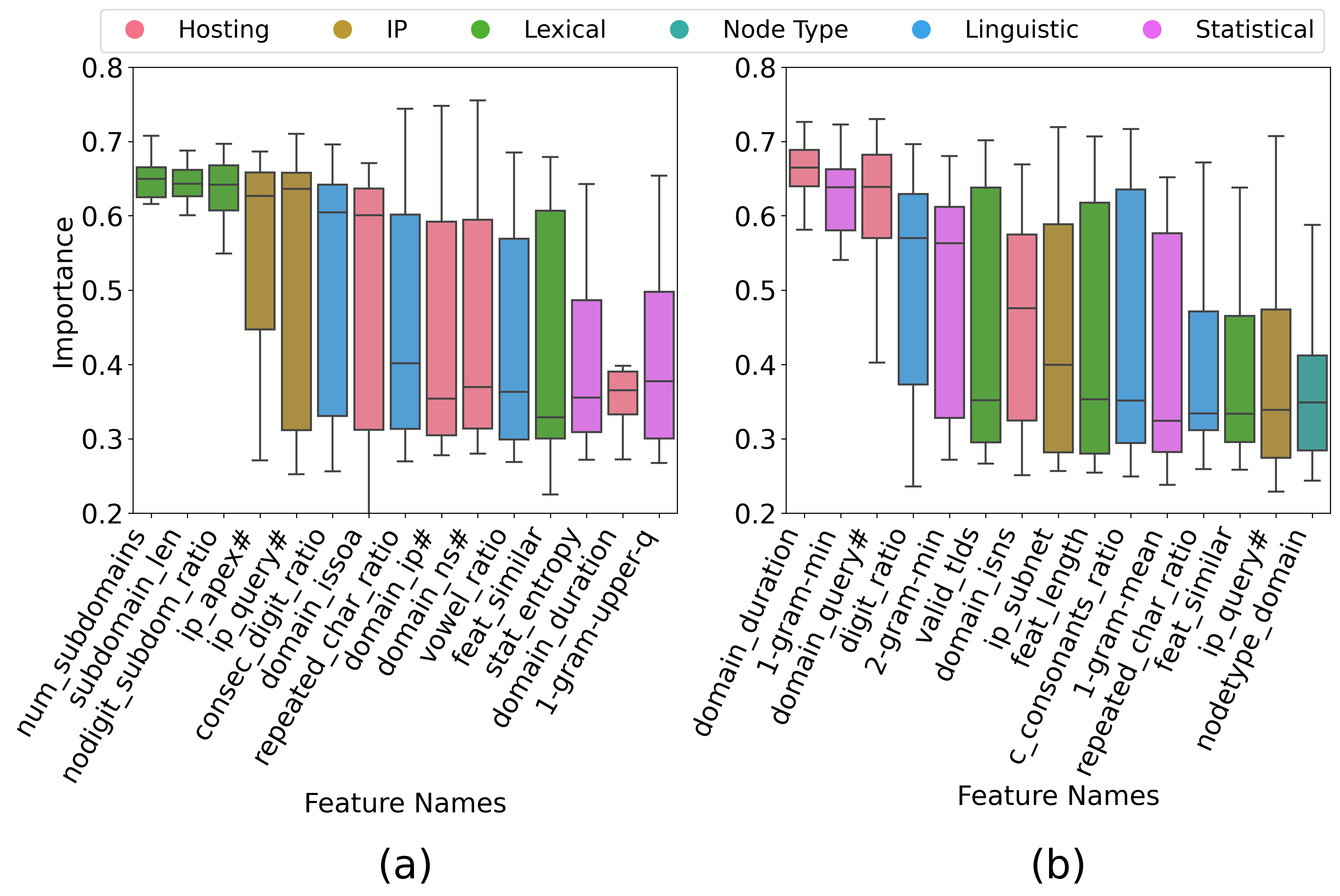}
  \caption{Feature importance for (a) false-positive and (b) false-negative predictions using perturbation-based method.}
  \label{fig:exp_explain_gnnexp_main}
\end{figure}

\spfinal{To enhance transparency and interpretability, we evaluated the importance of features and connections using two GNN explainers: a perturbation-based approach~\cite{ying2019gnnexplainer} and a gradient-based method~\cite{kokhlikyan:captum:2020}.
We categorized features into groups such as domain hosting, IP hosting, lexical, linguistic, and statistical, and assessed their significance across various prediction outcomes. Our analysis highlights the importance of hosting features (domain and IP) and certain lexical attributes, such as suspicious brands in domain names and matching nameservers, and the unimportance of features like length and the number of minuses. While the latter features can be easily circumvented by attackers, former ones, like longer durations with high query counts, would be hard for attackers to manipulate, thus increasing our robustness. We pay special attention to misclassifications and, as shown in Fig.~\ref{fig:exp_explain_gnnexp_main}, longer hosting durations and statistical lexical features contribute to false negatives, whereas lexical features have the greatest impact on false positives. Incorporating traffic pattern related hosting features and/or domain registration features likely to reduce such false positives.
Additionally, our analysis of the important connections within the expanded graph sheds light on potential malicious campaigns like the ones presented in Section~\ref{ss:campaigns}.
}

\subsection{Analysis of Detected Malicious Domains}
\label{subsec:analysisofdetectedmalicious}

We analyzed and validated a sample of malicious domains predicted (i.e., not part of the ground truth) by \system.
This post-analysis process includes rule-based daily status checking of a random set of daily malicious predictions for a predefined period. The pipeline is completed with a domain expert's manual verification of the remaining suspicious domains. In this way, we not only rely on the output of the machine learning models but also have the chance to observe new attack behaviors and update our approach.
For each day, we report VT malicious counts\revision{, GSB status} of the selected domains and active DNS (ADNS) resolutions \fatih{that represent non-NX and non-sinkhole domain counts}. 
Although \system is content-agnostic, we collect and check website contents to investigate additional signs of suspicion.
Our manual verification process is outlined in Appendix~\ref{app:sanity_table}.

\begin{figure}[t]
\centering
\includegraphics[width=0.85\columnwidth]{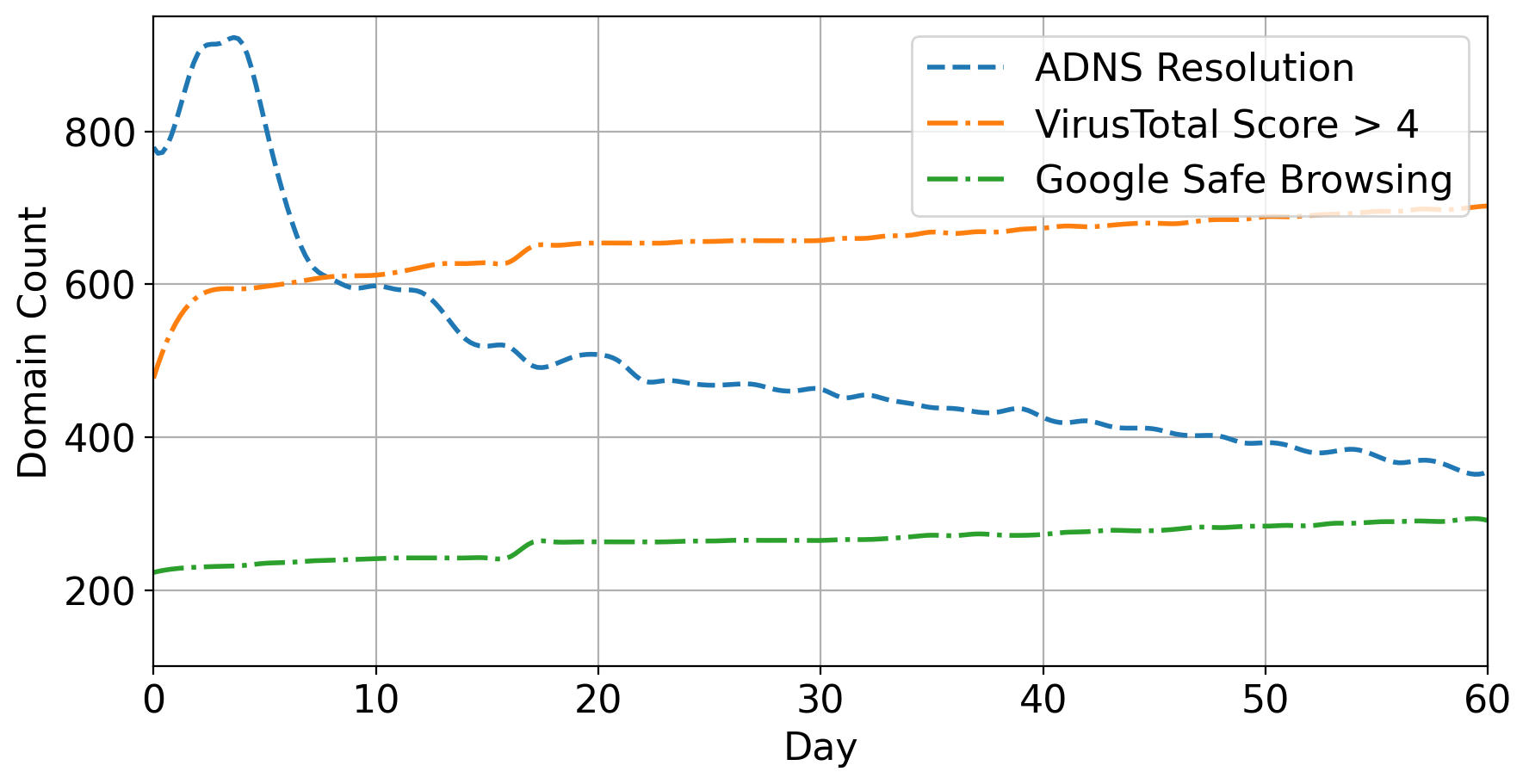}
\caption{Sanity checking for 60 consec. days. \eat{\nabeel{Let's add GSB results as well.}}}
\label{fig:sanity1}
\end{figure}

For randomly selected 1000 malicious predictions by \system on September 8, 2022, Fig.~\ref{fig:sanity1} shows the cumulative distribution of the detected domains by VT, \revision{GSB}, and ADNS resolution counts over time. \fatih{Upon the initial scan, the first-seen times within the collected reports reveal that 621 of these domains were originally submitted to VT by \system. This demonstrates the proactiveness of our discovery of malicious domains through passive DNS expansion. Among these 621 \system-submitted domains, $28.9\%$ and, for the cumulative one-thousand domains, $47.7\%$ of domains were initially classified as malicious by at least five engines. After two months, these figures increased to $64.6\%$ and $70.3\%$ for the \system-submitted domains and the overall. }

{\bf Active DNS Analysis:}
\fatih{ADNS resolutions provide valuable information for assessing the operational status of domains, offering insights into potential malicious activities. Upon classification, we note that 77.9\% of the selected domains immediately resolve to an IP address. Remarkably, within a span of just four days, this ratio increases to $91.5\%$, indicating a rapid deployment and activation of the identified domains.
After two months, this ratio decreases to $36.2\%$. This substantial decrease suggests a dynamic landscape of malicious activities.
These findings highlight the effectiveness of our GNN-based classifier in detecting malicious domains at an early stage, even before the web page contents are fully up and running. By leveraging passive DNS resolutions, \system demonstrates its capability to identify and flag potentially malicious domains during the early stages of their deployment.}

{\bf VT Status \& Content Analysis:}
After actively scanning the status of these domains for 60 days, we observe that $82.3\%$ of the domains are marked as malicious by at least one of the VT engines. Notably, none of the VT engines report $17.7\%$ of the domains as malicious, which is consistent with the findings of prior studies that report a range of $14.1\%$ to $18.2\%$ of undetected malicious domains~\cite{vissers:euric:2019, vissers:30day:2017}. It is concerning to note that existing approaches have a blind spot for these malicious domains. We attribute this to various reasons: not reported by a user, unavailability or insufficiency of the content when the scanning is performed, and domain cloaking. 
For the domains that do not have IOCs, after 60 days, we observe attributes akin to malicious domains: $54.8\%$ do not resolve to an IP, $48.6\%$ do not have any PDNS trace, $92.7\%$ have had less than 100 queries during the whole second month of the verification process, and none of them are present in the Internet Wayback Machine.
Regarding the ones that resolve to an IP, we observe that $6.3\%$ are parked, and $58.8\%$ either do not provide any HTML response, show empty content or various error messages such as invalid common name within the certificate, registration confirmation errors, or hosting provider-related warnings rather than legitimate content. 
Considering the attacker's hit and run strategy, in which malicious domains remain operational for a short duration, often just a single day in $60\%$ of cases~\cite{hao:30day:2013, oest:sunrise:usenix:2020}, our post analysis results comply with the expected attacker behavior.
\spupdate{Still, these are not concluding evidences for the maliciousness of domains. Many expired benign domains are also parked, and not all benign domains appear in the Wayback Machine. Nonetheless, these factors do raise suspicion regarding these domains.}
Table~\ref{tab:sanity_60days} in Appendix~\ref{app:sanity_table} further details the inspection results for different VT scores.

{\bf Precision and FPR:}
\fatih{Domains that resolve to an IP address, have HTTP content, are not marked as malicious by at least five VT engines or GSB are verified by a domain expert following the steps outlined in Appendix~\ref{app:sanity_table}. 
Through this rigorous process, we discover that 12 domains previously labeled as potentially malicious are, in fact, false positives. These domains and their details can be found in Table~\ref{tab:falsepositives} in Appendix~\ref{app:sanity_table}, demonstrating the precision of our sanity checking at an impressive 0.988.}
\fatihh{Despite our graph's relatively higher toxicity, benign predictions still dominate with a benign-to-malicious ratio of $\geq 10$. Consequently, the FPR ($\frac{FP}{FP + TN}$), remains impressively low ($\frac{12}{10,000}\approx0.1$\%), underscoring the exceptional efficacy of \system and aligning with the measured FPR using the ground truth we have collected.}

{\bf Comparison with GSB:}
\revision{We further compare \system with GSB. Specifically, we randomly pick 1000 predicted domains by \system and check how many are also marked as malicious by GSB. For those not marked by GSB, we followed the manual process described above conducted by a domain expert to check their status daily. We run this experiment over 6 different days. On average, around 18\% of those predicted by \system are marked by GSB on the first day. For the remaining domains predicted by \system, within the following two months, only an additional 12\% are marked by GSB as malicious within two months. This experiment shows \system's proactiveness, which can predict malicious domains much earlier than GSB, and \system's capability of detecting malicious domains missed by GSB. }

\spfinal{

\subsection{ Registration, Resolution \& Detection Times}~\label{ss:registrationdetection_analysis}

We also analyze the registration and PDNS first appearance times of the blocklisted domains to demonstrate the proactiveness of \system. Similar to prior work~\cite{hao:30day:2013, vissers:30day:2017}, we observe that the vast majority of malicious behavior of malicious domains occurs within 30 days after their registration. Specifically, $75.8\%$ of the predicted malicious domains are registered within the past month, and $96.1\%$ of the registrations occurred within the past year before detection. 
Our analysis further shows that benign domains take an average of 10 days from registration to their first PDNS appearance, while malicious domains appear in just 2 days. For newly discovered malicious domains registered in the past month, our system detects them on average within 3 days and 17 hours after their first PDNS appearance, compared to over 8 days for VT. This delay from VT is likely due to the reliance on submissions and the unavailability of domain contents. These statistics highlight the potential of our framework for earlier detection compared to traditional methods, which is quite valuable to registrars and hosting providers to detect abuses of their services early and take corrective actions.

\subsection{False Positive/Negative Analysis}~\label{ss:fpfn_analysis}

From the daily detection, we take random samples of 100 domains to perform error analysis periodically. We observed that the false positive rate is within the desired set threshold and the false positives are influenced by linguistic and statistical features. Specifically, domains containing consecutive numbers and minus characters were more likely to be flagged as malicious. As shown in Figure~\ref{sec5:feature_importance}, domains with no subdomains (just a landing page) or very long subdomains  also contributed to false positive predictions. Sample of such domains and their details can be found in Table~\ref{tab:falsepositives} in Appendix~\ref{app:sanity_table}. One may avoid such false positives by incorporating additional features such as WHOIS registration and TLS certificate features. In contrast, false negatives are associated with longer duration, high query counts, absence of suspicious brands in the domain name, presence of a matching name-server domain name, and low digit and consecutive consonant ratios—traits more commonly associated with benign domains. Content based features such presence of forms, brand logos, and injected scripts may assist in reducing such false negatives. 
}

\subsection{Example Campaigns Detected}~\label{ss:campaigns}
\begin{figure}
\centering
\begin{subfigure}[t]{0.85\linewidth}
    \includegraphics[width=\linewidth]{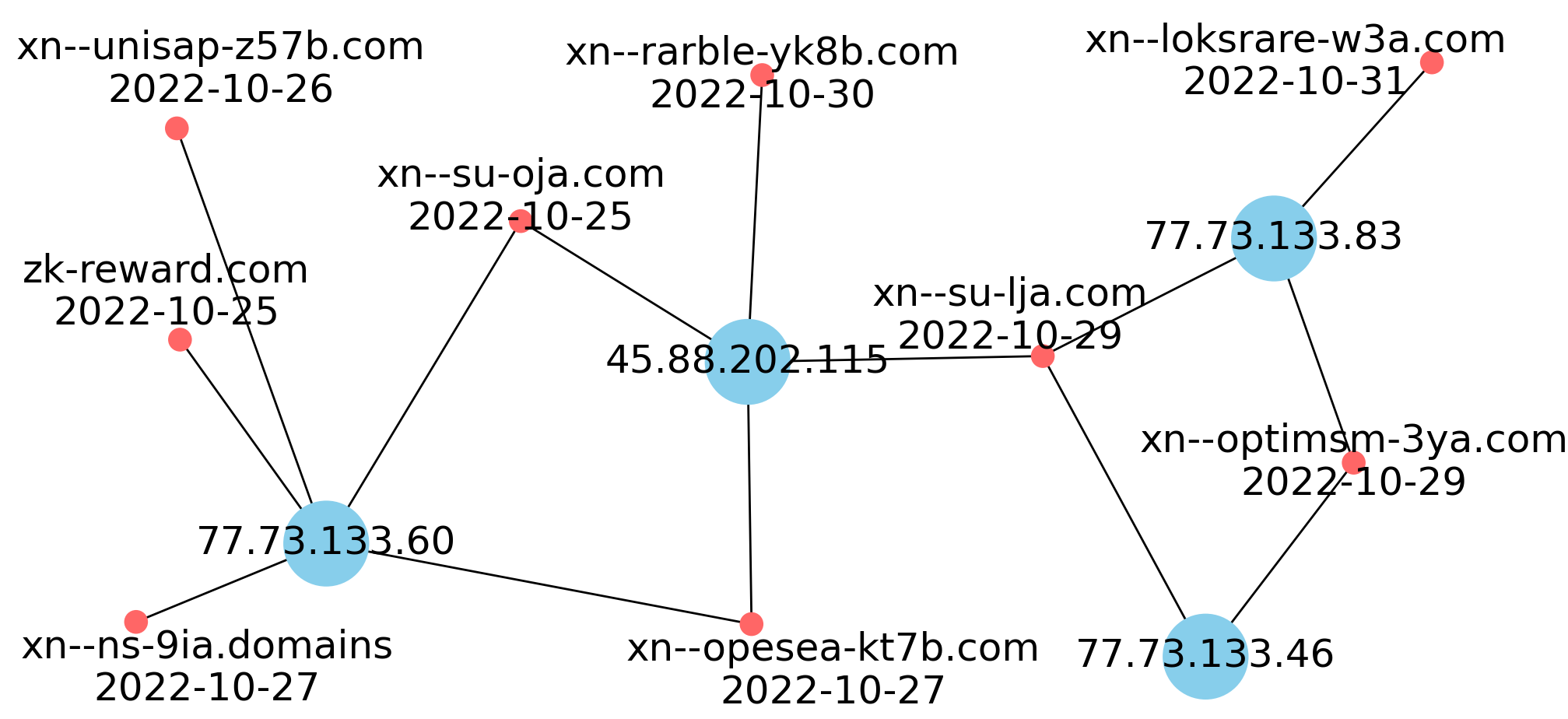}
    \caption{}
    \label{fig:campaignweb3}
  \end{subfigure}  \hfill
  \begin{subfigure}[t]{0.85\linewidth}
    \includegraphics[width=\linewidth]{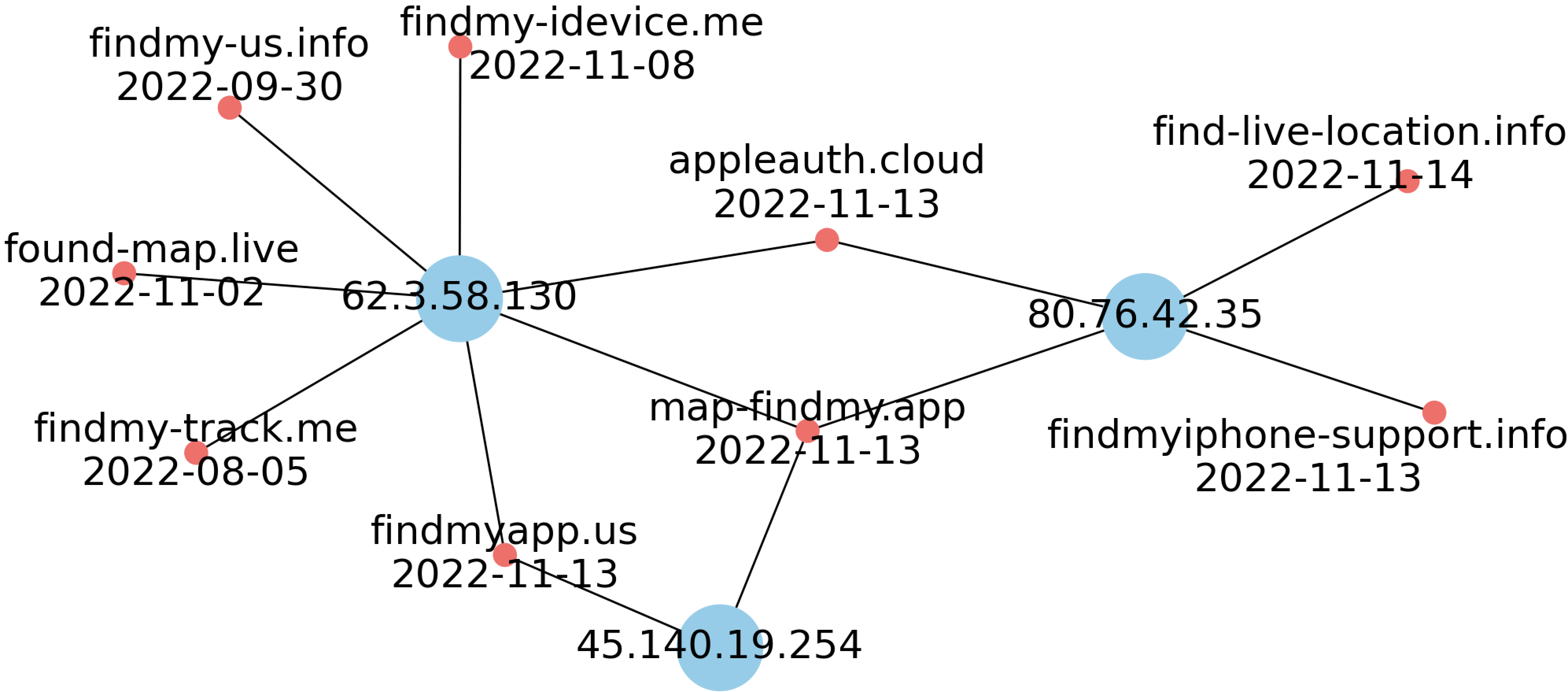}
    \caption{}
    \label{fig:campaignicloud}
  \end{subfigure}\hfill
  \caption{Campaigns abusing (a) web3 domains (b) Apple/iCloud, along with resolved IPs and first detection dates.}
  \label{fig:campaigns}
\end{figure}

In this section, we investigate a couple of example campaigns detected by \system. 
Fig.~\ref{fig:campaignweb3} shows a campaign where web3 domains such as \url{opensea.com}, \url{ens.domains}, and \url{rarible.com} are abused. The threat actor predominantly uses homographs to trick users. 
An interesting observation is that the campaign hosts incrementally more domains each day. However, all of them are hosted within one day of registration. Fig.~\ref{fig:campaignicloud} shows another campaign that abuses Apple and/or iCloud. \system starts to detect the first of the malicious domains and continues to crawl the same infrastructure to proactively discover additional malicious domains as and when they are hosted. Unlike the web3 campaign, we observe that the threat actor not only hosts domains over a long period of time but also strategically waits after domain registration, possibly to avoid detection by domain reputation systems. For example, \url{map-findmy.app} was created on 2022-10-02, but used to launch an attack on 2022-11-13.

\section{Related Work}~\label{sec:related}
{\bf Early Detection.} PhishNet~\cite{phishnet:2010} proposes five
heuristics to enumerate simple combinations of known phishing sites to discover new phishing URLs. Felegyhazi et al.~\cite{predictive:leet:2010} identify new malicious domains leveraging DNS zone files and WHOIS records. They observe that malicious domains are registered in bulk and are utilizing either fresh or self-resolving name servers. Based on these observations, they cluster domains and identify domains not discovered by existing blocklists. EvilSeed~\cite{evilseed:sp:2012} proposes an approach to identify likely malicious webpages based on the links in seed malicious pages.  Predator~\cite{Predator_Hao2016} attempts to identify malicious domains at the time of registration. Their intuition is that spam domains are registered in bursts and often at similar stages in the domain life cycle. They rely on the WHOIS records at the domain registration, historical WHOIS records for domains, and the WHOIS records of domains registered at the same registrar. While these approaches detect malicious domains that are lexically similar or registered in bursts, we remark that the detected set is only a small subset of all malicious domains out there as many malicious domains do not exhibit lexical similarity nor time-based correlation.

{\bf Feature Based Classification.}
Many approaches~\cite{maldom:kdd:2009,Fukuda:2015:DNSBackScatter,Segugio_Rahbarinia2015,DeepDGA_Anderson2016,MethodForDetectingDgaBotnet_Tong2016}, including Notos~\cite{Notos_Antonakakis2010} and EXPOSURE~\cite{bilge:2014:Exposure}, identify malicious domains by building a classifier using the local features extracted from passive DNS data along with other network information such as WHOIS records~\cite{Liu:2015:CLP}. Such approaches are effective as long as the local features used in the classification are not manipulated. However, it has been shown~\cite{TowardsSystematicEvaluationOfTheEvadabilityOfBotnetDetectionMethods_Stinson2008} that many local features such as TTL-based features and patterns in domain names, are easy to manipulate and thus rendering such techniques less effective. These approaches perform best when one has access to sensitive individual DNS queries which are difficult to gain access to. On the other hand, graph-based approaches like ours can detect malicious domains with high accuracy using only aggregate DNS data which is relatively easier to gain access to. 
{\bf Graph Based Detection.}
Nabeel et. al.~\cite{bp_mal2:2020} utilize the passive DNS data and graph inference algorithms such as belief propagation and path-based algorithms for malicious domain defections. 
A recent study by Kim et. al.~\cite{BPPhishingCCS:2022} uses loopy belief propagation with adaptive edge potentials on a heterogeneous network consisting of URLs, domains, IPs, and word segments. \revision{Simeonovski et al.~\cite{internetgraph2017} propose a taint-style propagation technique based on a set of rules that focuses on a graph structure around the top 100K Alexa domains.} 
MalRank~\cite{malrank:acsac:2019} implements a large-scale graph inference algorithm to detect malicious domains in SIEM environments. 
HinDom~\cite{hindom:raid:2019} implements a malicious domain detection system by representing DNS in a heterogeneous information network (HIN) and a meta-path based path-similarity classifier.  Meanwhile, HGDom~\cite{hgdom:NOMS:2020} utilizes a heterogeneous variant of GCN called MAGCN for domain classification. Ringer~\cite{ringer:ICCS:2020} proposes a scalable method to detect malicious domains by a dynamic graph neural network. DeepDom~\cite{deepdom:cns:2020} uses SHetGCN to classify malicious domains, a heterogeneous variant of GCN. The graph contains clients, domains, IP addresses, accounts, and relationships in between. GAMD~\cite{ahgnn}, Blocklist-Forecast~\cite{blocklist_raid2024} and HANDom~\cite{handom:wang:2023} utilize variations of heterogeneous graph neural networks for malicious domain prediction. The DNS graph contains clients, domains, IPs, and relationships among them. 
\eat{80\% of their dataset is based on a dataset from 2017 that may not reflect current attack behaviors. Their precision values are also quite low compared to our approach (0.832), which can restrict their usage in real-life.}

Though the above models show promising results, they have limitations when predicting at a global scale:(1) belief propagation based approaches fail to take advantage of features associated with domains whereas our approach leverages both node features and network topology; (2) existing approaches require user-level DNS query, which is often limited, to build the network graph whereas our approach utilizes publicly available aggregated DNS resolution information; (3) They are not able to predict on unseen domains without retraining from the scratch, whereas our approach supports inductive learning without retraining; and (4) They use static graphs and do not show if they generalize to unseen data, whereas our approach is generalizable over time.

\section{Limitations}~\label{sec:limiations}
PDNS has visibility to around 90\% of the domains on the Internet. This, in turn, results in a reduction of 10\% of seed domains from our ground truth. We attribute this limitation to the data source we utilize. One may utilize additional passive DNS sources such as Spamhaus~\cite{Spamhaus}, circl.lu~\cite{Circl}, or Rapid7~\cite{Rapid7}, to augment Farsight PDNS coverage. Further, an active DNS lookup from multiple vantage points may assist further improve the coverage.

Our approach is not effective at detecting malicious websites created on webhosting services. The key reason is that these sites are hosted on benign infrastructures and do not exhibit homophily relationships. One may construct either a different model such as a content-based classifier to detect such domains. It is fairly normal practice that many detectors are deployed to detect different types of malicious websites such as attack domains, malicious webhosting domains, and DGA domains. 
The graph-based detection mechanism is unable to detect compromised domains unless content-based features are incorporated. As compromised domains are benign domains that turned malicious, their hosting neighborhood has very low toxicity and the neighboring domains in DNS graph are more likely to be benign. Further, features of compromised domains are quite similar to those of benign domains. Thus, an alternative approach is required to detect compromised domains. There are recent research efforts to detect compromised domains~\cite{comp_or_at:2021:usenix,comar:esp:2020}, which complement ours. \spfinal{Further, based on the observation that compromised domains are often infected with malicious scripts such as miners and skimmers, and/or redirect to low reputed URLs, one may construct a graph leveraging content based relationships and utilize a similar graph learning approach to identify compromised domains.}

\section{Conclusions}~\label{sec:conclusion}
We build \system, a highly accurate malicious domain detection system that has been operational since June 1st, 2022 to the present. We leverage the observation that attackers reuse hosting infrastructures to host their disposable domains to detect malicious domains based on a small set of known malicious domains. Our models detect, on average, 19K malicious domains per day, which is 5 times the number of newly observed malicious domains in VT. Since our PDNS crawler proactively identifies domains in the neighborhood of malicious seed domains, \system detects malicious domains way ahead of VirusTotal and GSB, often several weeks ahead. Our sanity checking reveals the concerning fact that VT and GSB have a blind spot for at least 20\% of the detected malicious domains even after multiple scanning. In order to improve the coverage of detection, our approach can be easily extended to incorporate other associations such as registration records, certificates, and redirection chains. An interesting future direction is to automatically identify attack campaigns.

\bibliographystyle{IEEEtran}
\bibliography{main}

\begin{appendices}

\section{Graph Toxicity}
~\label{app:graphtoxicity}

\begin{figure}[ht]
\centering
\includegraphics[width=0.90\columnwidth]{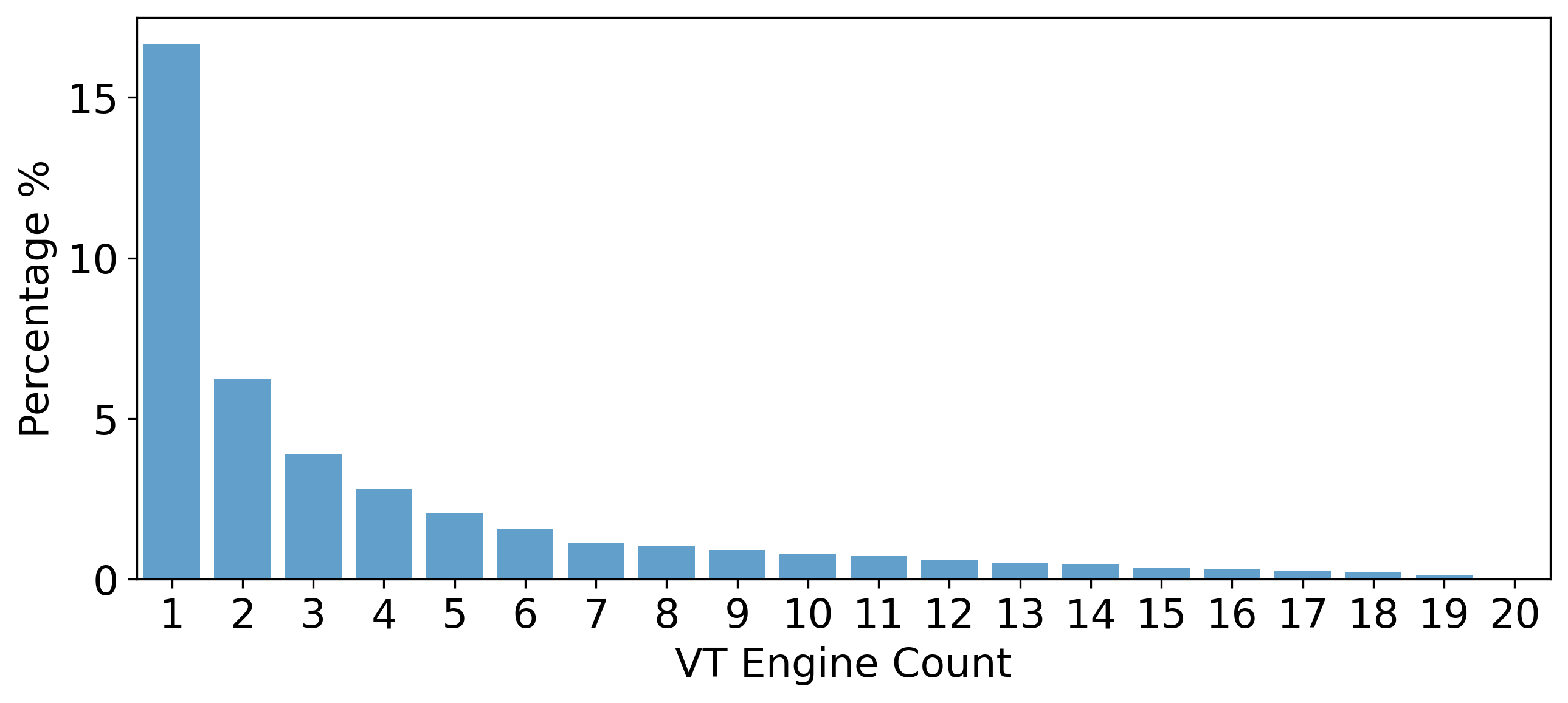}
\caption{\ccsupdate{VT Positives Distribution of Seed Domains' Neighbors. The $y$-axis indicates the percentage of nodes flagged as malicious by at least $x$ VT engines. Notably, 16.65\% of previously unknown domains are flagged as malicious by at least one engine, indicating our success in reaching highly malicious neighborhood.}}
\label{fig:random_vt}
\end{figure}

A key conceptual novelty in our work is the automated construction of a graph around attack domains and the guided graph expansion to create a graph with high toxicity. 
To measure the difference in toxicity, we take a daily first seen malicious domains from VT and construct a graph around these domains using passive DNS and check the VT scores of the domains in the expanded graph.
We then assess the VT scores associated with the domains within the expanded graph and compare this with the VT scores of a random sample from passive DNS using the threshold VT $\geq$ 5. The distribution of VT positives in a randomly sampled expanded graph is depicted in Fig.~\ref{fig:random_vt}. This experiment complements the results presented in Fig.~\ref{fig:overlapping_ip_july}. For that experiment, we collected the IP resolutions (using passive DNS) of the malicious domains (VT $\geq$ 5) and checked how many IPs were previously reported as hosting a malicious domain during the past week. 

\section{Node Features}
\label{app:appnodefeatures}

\review{
In our pursuit of creating a practical system, we purposefully selected feature sets that are widely accessible and commonly employed in the cybersecurity field. Specifically, we adapted or derived several features used in a similar domain from various sources and combined them with our novel features, as shown in Table~\ref{tab:features}. Notably, we made a conscious choice to omit content-related features and those necessitating premium access, such as those from VirusTotal. This decision was driven by our pursuit of a more practical approach, which is scalable and computationally efficient.
In our research, we utilize Farsight data as our source for PDNS. A key advantage of PDNS is its capability to safeguard the privacy of individual Internet users, as it exclusively contains aggregated data. We harnessed the PDNS repository to expand around seed malicious domains and to gather domain/IP features for our study.
}

\begin{table}[htbp]
\caption{Node Features}
\label{tab:features}
\centering
\resizebox{\linewidth}{!}{
\begin{tabular}{|p{3cm}|p{4cm}|c|}
\hline
\textbf{Feature Name} & \textbf{Description} & \textbf{Derived from} \\ \hline
\multicolumn{3}{|c|}{\textbf{Domain Lexical Features}} \\ \hline
pop\_keywords & Suspicious popular keywords count & \cite{Kintis:2017:Combosquatting}\\ \hline
length & Length of domain name & \cite{lexical2015,page:2019:mal, practicalattacks:SP:2024}\\ \hline
minus & Number of minus signs in domain name & \cite{lexical2015,page:2019:mal, Predator_Hao2016}\\ \hline
suspicious\_tld & Presence of suspicious TLD & \cite{phishingcatcher}\\ \hline
brand\_pos & Position of brand in domain name & New \\ \hline
similar & Presence of term resembling recognized brand & New \\ \hline
fake\_tld & Number of gTLDs in domain name & New \\ \hline
num\_subdomains & Number of subdomains & \cite{phicious:raid:2022, Notos_Antonakakis2010, practicalattacks:SP:2024}\\ \hline
subdomain\_len & Mean subdomain length & \cite{Notos_Antonakakis2010, ringer:ICCS:2020}\\ \hline
has\_www & Presence of www prefix & \cite{ringer:ICCS:2020}\\ \hline
valid\_tlds & Presence of valid TLD & \cite{ringer:ICCS:2020}\\ \hline
has\_single\_subdomain & Presence of single-character subdomain &  \cite{ringer:ICCS:2020}\\ \hline
has\_tld\_subdomain & Presence of TLD as subdomain & \cite{ringer:ICCS:2020}\\ \hline
digit\_ex\_subdomains\_ratio & Ratio of digit-exclusive subdomains & \cite{ringer:ICCS:2020}\\ \hline
has\_ip & Presence of IP address & \cite{ringer:ICCS:2020}\\ \hline
\multicolumn{3}{|c|}{\textbf{Domain Hosting Features}} \\ \hline
query\_count & Access count in last 30 days & \cite{phicious:raid:2022, Exposure_Bilge2011}\\ \hline
\#ips & Number of IPs hosting domain & \cite{phicious:raid:2022, practicalattacks:SP:2024}\\ \hline
\#name\_servers & Number of authoritative name servers & \cite{phicious:raid:2022, comp_or_at:2021:usenix}\\ \hline
is\_ns\_matching & Matching apex with name server & \cite{phicious:raid:2022}\\ \hline
\#soa\_domains & Number of SOA domains & \cite{comp_or_at:2021:usenix}\\ \hline
is\_soa\_matching & Matching apex with SOA domain & \cite{comp_or_at:2021:usenix}\\ \hline
duration & PDNS duration of domain & \cite{phicious:raid:2022, Exposure_Bilge2011}\\ \hline
\multicolumn{3}{|c|}{\textbf{IP Features}} \\ \hline
\#apexes & Number of apex domains on IP & New \\ \hline
query\_count & Access count of hosted domains & New \\ \hline
duration & PDNS duration of IP & New \\ \hline
subnet & Encoded class C subnets & New \\ \hline
asn & Autonomous System Number & New \\ \hline
\multicolumn{3}{|c|}{\textbf{Statistical \& Linguistic Features}} \\ \hline
entropy & Domain name entropy & \cite{comar:esp:2020, phicious:raid:2022, practicalattacks:SP:2024}\\ \hline
N\-gram (N = 1, 2, 3) & Mean, median, and std dev of N\-grams &  \cite{Notos_Antonakakis2010, ringer:ICCS:2020}\\ \hline
underscore\_ratio & Ratio of underscores & \cite{phicious:raid:2022, ringer:ICCS:2020}\\ \hline
has\_digits & Presence of digits & \cite{Exposure_Bilge2011, phicious:raid:2022, ringer:ICCS:2020}\\ \hline
digit\_ratio & Digit ratio & \cite{Exposure_Bilge2011, phicious:raid:2022, ringer:ICCS:2020}\\ \hline
vowel\_ratio & Vowel ratio & \cite{ringer:ICCS:2020}\\ \hline
alphabet\_cardinality & Alphabet cardinality & \cite{ringer:ICCS:2020}\\ \hline
repeated\_char\_ratio & Ratio of repeated characters & \cite{ringer:ICCS:2020}\\ \hline
consec\_consonants\_ratio & Ratio of consecutive consonant pairs &  \cite{ringer:ICCS:2020}\\ \hline
\end{tabular}
}
\end{table}

\begin{table*}
\caption{False Positive Domains Detected in Sanity Checking}
\label{tab:falsepositives}
\centering
\resizebox{\linewidth}{!}{
\begin{tabular}{|c|c|c|c|c|c|c|c|c|c|c|c|}
\hline
Domain Name & Freenom & Brand Squatting & ADNS Resolves & Parking & Content Length & VT Pos. & Registrar & Registration Date & PDNS Duration & \#PDNS Query \\
\hline
wowerides.ca & False & False & True & False & 730712.0 & 0 & go get canada domain registrar ltd. & 2022-02-10 & 175.355509 & 10 \\
biopell-medical.com & False & False & True & False & 466277.0 & 0 & hosting ukraine llc & 2022-08-22 & 155.806412 & 71  \\
apamall.it & False & False & True & False & 67855.0 & 0 & NaN & 2022-07-08 & 128.187535 & 52  \\
univers-sabeauty-wellness.com & False & False & True & False & 125200.0 & 0 & realtime register b.v. & 2022-08-30 & 156.943808 & 117 \\
blckwave.com & False & False & True & False & 57048.0 & 0 & hosting ukraine llc & 2022-09-02 & 171.490370 & 96  \\
the-paddock.be & False & False & True & False & 103291.0 & 0 & NaN & NaN & 171.832072 & 101  \\
scholarcy.ai & False & False & True & False & 449748.0 & 0 & namecheap & 2021-06-14 & 33.240637 & 6  \\
novanclinic.com & False & False & True & False & 189757.0 & 0 & namecheap inc & 2022-08-22 & 156.773738 & 177  \\
elbidondeclaudia.online & False & False & True & False & 184873.0 & 0 & hostinger, uab & 2022-06-28 & 172.093970 & 172  \\
lunatic-studio.com & False & False & True & False & 358586.0 & 0 & hosting ukraine llc & 2022-08-25 & 146.747535 & 59  \\
xn--podbitka-chem-7hc.pl & False & False & True & False & 60371.0 & 0 & NaN & NaN & 142.765336 & 60  \\
lasyshark.shop & False & False & True & False & 260915.0 & 0 & namecheap, inc. & 2022-07-28 & 140.863160 & 44 \\
\hline
\end{tabular}}
\end{table*}

\section{Grid Search \& Computational Performance}
\label{app:differentgnn}

\begin{figure}[ht]
\centering
\includegraphics[width=1.0\columnwidth]{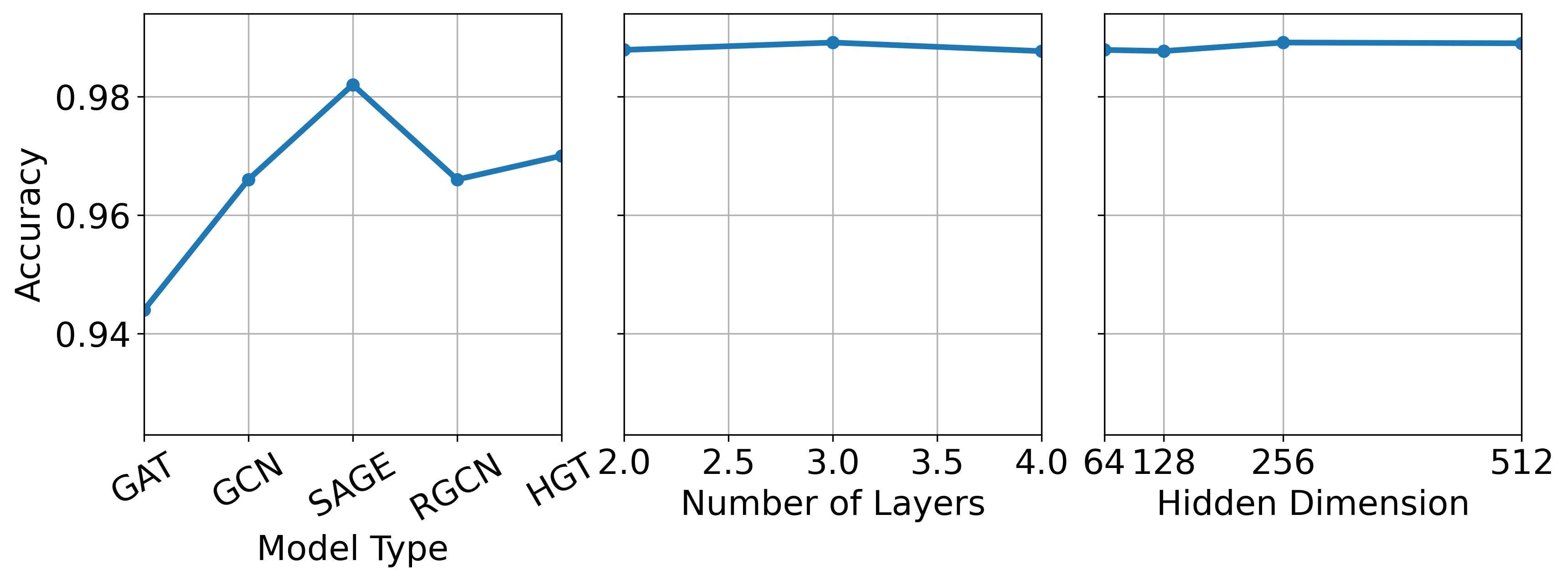}
\caption{Impact of GNN architectures, embedding dimensions, and layer count.}
\label{fig:grid_search}
\end{figure}

After assessing various semi-supervised GNN models for malicious domain classification, we identify GraphSAGE as the optimal choice based on our grid-search analysis, as depicted in Fig.~\ref{fig:grid_search}. GraphSAGE outperforms others in terms of classification accuracy, precision, and FPR, making it ideal for subsequent experiments and daily pipelines. Based on the grid search results we use GNN that comprises three layers with embedding dimensions of 256, employ neighbor sampling, and a final layer that aggregates all embeddings from the preceding layers. 

A key goal of our research is to make \system a practical system; thus, we conducted a comprehensive assessment of its computational performance. In our testbed, we utilized a 16-core Intel Xeon processor server with 32 GB of memory running CentOS Linux for data collection and processing and Tesla V100 GPUs for model training. Within this setup, the daily blocklist generation, including graph construction, feature and ground truth collection, model training, and prediction, takes only 2 hours to complete. Additionally, on-demand graph construction, feature collection, and predictions exhibit an average response time of 8 seconds without load balancing. These performance metrics demonstrate the efficiency and feasibility of \system in real-world scenarios. Considering that around 120K \textit{.com} domains are registered per day and \textit{.com} constitutes nearly 50\% of all registered domains, this translates to less than 2 domains per second. A naive approach to scaling would be to have just 16 instances to handle the load, which is a relatively smaller number of instances compared to the industry deployment involving 100s of instances.

\begin{figure}
\centering
\includegraphics[width=0.75\columnwidth]{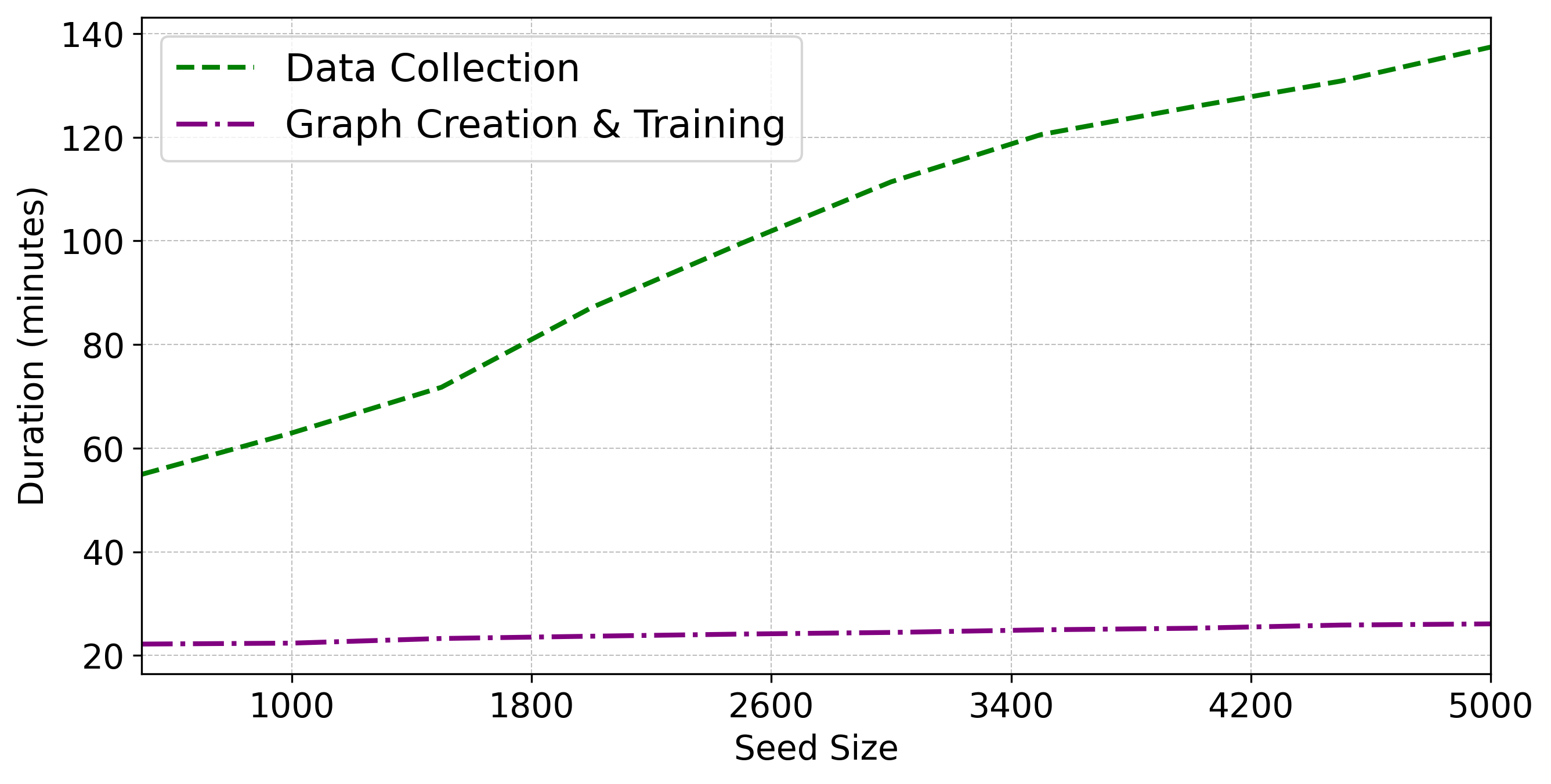}
\caption{Daily blocklist generation time by seed size.}
\label{fig:comp_perf}
\end{figure}

In Fig.~\ref{fig:comp_perf}, we evaluate the time required to generate the blocklist based on variable seed counts. We observe that the time required for the data collection step increases proportionally with the seed size. This cost could be reduced by utilizing a distributed PDNS database, which currently is a single instance.
\eat{It should be noted that , as we utilize a locally hosted PDNS database, the expansion and feature collection process can be distributed across multiple machines, enabling further performance improvement.} Additionally, we found that the graph generation and model training steps combined exhibit minimal sensitivity to the seed size, further emphasizing the system's scalability.

\section{Quality Checks \& Post Analysis}
\label{app:sanity_table}

\spfinal{Regular assessments of our benign ground truth indicate the quality of these domains. At the end of each month, we execute the sanity-checking pipeline discussed in Section~\ref{subsec:analysisofdetectedmalicious} on the benign ground truth. This involves checking the VT status, certificates, and content of the domains. We perform manual checks for domains flagged by at least one of the VT engines. 
During manual verification, we investigate each domain in terms of the presence of phishing and fraudulent activities, distribution of malware, malicious or harmful content, and involvement in different types of brand squatting attacks. To make the manual assessment more efficient, we automatically collect several types of auxiliary information of domains: historical registration records, TLS certificate, screenshots, HTML content, PDNS records, active DNS records, latest VT report, underlying subgraph around the domain, and previous assessments on the domain. We investigate the website content, check the Internet Wayback Machine snapshots, and evaluate the reports from two threat intelligence platforms: Microsoft Defender Threat Intelligence (ti.defender.microsoft.com) and AlienVault (otx.alienvault.com). With all this information available, a domain expert, on average, spends 2-3 minutes to manually verify.
Using this pipeline, we only needed to evaluate less than $2\%$ of our ground truth and relabel a small fraction of domains corresponding to less than $0.1\%$ of the benign ground truth. }

\begin{table}
\centering
\caption{Sanity Checking Results After 60 Days.}
\resizebox{\linewidth}{!}{
\begin{tabular}{|l|c|c|c||c|}
\hline
& \multicolumn{4}{c|}{Domain Count} \\ 
\cline{2-5} 
    & VT score=0 & 0$<$VT score$<$5 & VT score$\ge$5 & Total \\
    \hline
NX Domain & 97 & 77 & 479 & 653   \\ \hline
No Content & 47 & 21 & 66 & 134    \\ \hline
Parked Domain & 5 & 3 & 6 & 14  \\ \hline
Brand Impersonating & 2 & 5 & 28 & 35    \\ \hline
Manual Verification & 26 & 14 & 124 & 164   \\ \hline
\hline
Total & 177 & 120 & 703 & 1000\\ \hline
\end{tabular}
}
\label{tab:sanity_60days}
\end{table}

As a testament to the quality checks we perform on the sample of generated daily blocklists.
Table~\ref{tab:sanity_60days} summarizes the inspection results for various VT score groups and Table~\ref{tab:falsepositives} provides a sample of false positive predictions with further details.
This data is based on a randomly selected 1000 malicious predictions made by \system.
Among the domains marked as safe by all the VT engines, we encounter various attack types, such as phishing domains (\url{americafirstsecr.com}, \url{app-2q3fob.com}), malware distribution sites (\url{nsupport360.cc}, \url{yarbiegishola.xyz}), brand squatting domains (\url{comptes-paypal.com}, \url{appletw.net}), DGA domains (\url{sgz25cr.cn}, \url{5yrkso9.us}), among others. 
These examples serve as representative cases illustrating the diverse range of detected malicious activities.
When we checked the Sophos category of malicious predictions, we observed that 86\% of the malicious predictions are in the ``phishing and fraud", 13\% are in the ``malware and spyware", and 1\% are in the ``command and control" category.

\newpage

\section{Meta-Review}
\subsection{Summary}
This paper introduces Mantis, a system designed to detect zero-day malicious domains by monitoring low-reputation hosting infrastructure.  The goal of the framework is to accurately classify domains at the time of hosting setup, but prior to the deployment of malicious content on the respective domains. Mantis uses a content-agnostic approach wherein network, IP, and lexical features are labeled with ground truth data from domain rankings, passive DNS, and malicious domain feeds. Mantis then employs Graph Neural Networks to analyze hosting patterns and predict malicious domains. A real-world deployment of the framework showed daily detections with high precision (99.7\%) and recall (86.9\%), and achieved a low false positive rate of 0.1\%.

\subsection{Scientific Contributions}
\begin{itemize}
\item Provides a New Data Set For Public Use
\item Creates a New Tool to Enable Future Science
\item Provides a Valuable Step Forward in an Established Field
\end{itemize}

\subsection{Reasons for Acceptance}
\begin{enumerate}

\item Proactive detection: MANTIS can predict malicious domains days to weeks before they appear on popular blocklists, highlighting its proactive nature.
\item Comprehensive evaluation: the paper performs a comprehensive set of experiments and conducts a baseline comparison with existing ML approaches and SoTA (Tables 4 and 5). The paper also reports the computational performance.
\item High accuracy: the reported precision, recall, low false positive rate, and adversarial robustness support practical applications.
\item Operational system: Mantis has been already operational for over a year, consistently detecting a significant number of malicious domains daily, which underscores its practicality.
\end{enumerate}

\subsection{Noteworthy Concerns}
\begin{enumerate}
\item Detection limitations: the focus on attacker-created domains might overlook the importance of detecting compromised domains, which are also significant in real-world scenarios. The paper acknowledges this and discusses potential solutions that are left for future work. Mantis is also not designed to distinguish between benign and malicious domains within certain shared hosting environments. Benign subdomains that share legitimate infrastructure with malicious ones would lead to false positives and are excluded from this work.
\item The detection performance of the system requires reliable ground truth data from large scale oracles. Novel attack vectors that differ significantly from the training data may not be detected. 
\item The practical potential of the proactive detection possible by the framework is estimated (i.e., detection at the time of hosting setup) but not empirically evaluated.
\end{enumerate}

\section{Response to the Meta-Review}

\begin{enumerate}
    \item Detection limitation: As we discuss in the Limitations Section, our approach augments the existing compromised domain detectors and rentable domain detectors. One may devise novel graph based approaches to improve existing compromised and/or rentable domain detectors.
    \item Large oracles: While GT from large oracles greatly improves the performance in terms of precision and recall, GT from small oracles such as PhishTank can still detect malicious domains with over 90\% precision and recall. Similar to other DL based approaches, if the attack vector is completely novel from the training data, it is likely to have a blind spot. We recommend retraining the model periodically to minimize such blind spots.
    \item Proactive detection: In Section 6.3, we show our approach is several days more proactive compared to VirusTotal, which is the most popular and the largest domain maliciousness lookup service. In order to further improve the proactiveness, one may execute our pipeline at intervals shorter than 1 day, for example, every 6 hours. We leave further empirical evaluation utilizing different windows and/or known malicious domains as future work.
\end{enumerate}
\end{appendices}

\end{document}